\documentclass[aps,prd,onecolumn,preprintnumbers,notitlepage,superscriptaddress,
nofootinbib,amsfont,amssymb,10pt,singlespacing] {revtex4-1}
\usepackage{amsmath,bm}
\usepackage{times}
\usepackage{braket}
\usepackage{color,graphicx}
\usepackage{slashed}
\usepackage{hyperref}
\usepackage{graphics}
\usepackage{subfigure}
\usepackage{multirow,makecell}
\usepackage{textcomp}
\usepackage{mathtools}
\usepackage{float}

\newcommand{\beq}{\begin{eqnarray}}
\newcommand{\eeq}{\end{eqnarray}}

\def\lsim{ {\ \lower-1.2pt\vbox{\hbox{\rlap{$<$}\lower6pt\vbox{\hbox{$\sim$}
}}}\ } }
\def\gsim{ {\ \lower-1.2pt\vbox{\hbox{\rlap{$>$}\lower6pt\vbox{\hbox{$\sim$}
}}}\ } }
\definecolor{Red}{rgb}{1.,0.,0.}

\definecolor{Blue}{rgb}{0.,0.,1.}

\definecolor{nicered}{rgb}{0.7,0.1,0.1}
\definecolor{nicegreen}{rgb}{0.1,0.5,0.1}

\hypersetup{colorlinks,citecolor=nicegreen,linkcolor=nicered}

\begin{document}

\title{Study of quasi-two-body $B^{0}\rightarrow T(\pi\pi, K\bar{K},\pi\eta)$ decays in perturbative QCD approach}

\author{Lun-Lian~Mu}
\email[Electronic address:]{mll2631977415@163.com}
\affiliation{School of Physical Science and Technology,
 Southwest University, Chongqing 400715, China}

\author{Xian-Qiao~Yu}
\email[Electronic address:]{yuxq@swu.edu.cn}
\affiliation{School of Physical Science and Technology,
Southwest University, Chongqing 400715, China}

\date{\today}

\begin{abstract}

In this study, we calculate the $CP$-averaged branching ratios and the direct $CP$-violating asymmetries of the three body decays $B^{0}\rightarrow T(\pi\pi, K\bar{K},\pi\eta)$ in the perturbative QCD approach, where $T$ denotes tensor mesons $a_{2}(1320)$, $K^{*}_{2}(1430)$, $f_{2}(1270)$ and $f^{'}_{2}(1525)$, and the daughter branching is $f_{0}(980)\rightarrow \pi\pi, K\bar{K}$,$f_{0}(500)\rightarrow \pi\pi$, $a_{0}(980)\rightarrow K\bar{K}, \pi\eta$. By introducing two-meson distribution amplitudes parametrized by the timelike form factors, the three-body decay is simplified to quasi-two-body decay. The scalar mesons $a_{0}(980)$, $f_{0}(980)$ and $f_{0}(500)$ are regarded as the lowest-lying $q\overline{q}$ state, considering the effect of mixing angle $\theta$ between $f_{0}(980)$ and $f_{0}(500)$, $\phi$ between $f_{2}(1270) $and $f^{'}_{2}(1525)$ on our calculations. We found the following (a) With $\theta=135^{\circ}$, we evaluate the branching fractions of $B^{0} \rightarrow K^{*}_{2}(1430)f_{0}(980)[ \rightarrow \pi^ {+}\pi^{-}]$ to be $4.30\times10^{-6}$,then under the narrow-width approximation we extract the branching fraction of the decay $B^{0} \rightarrow K^{*}_{2}(1430)f_{0}(980)$ to be $8.78\times10^{-6}$,which is consistent with the current experimental data well. (b) The decay rates for the considered decay modes are generally in the order of $10^{-8}$ to $10^{-5}$. (c) The branching fractions are sensitive to the $\theta$, and the opposite is true for $\phi$. The $\phi$ is really small, so the decay branching ratio has only little change, except for some decays involving $f^{'}_{2}(1525)$. (d) The $\theta$ and $\phi$ can bring remarkable change to the direct $CP$ asymmetries of pure penguin processes so that it is not 0. (e) We calculate the relative partial decay widths ${\Gamma}(a_{0} \rightarrow K\overline{K})/{\Gamma}(a_{0} \rightarrow \pi\eta)$ and the ratio ${\cal B}(f_{0} \rightarrow K^{+}K^{-})/{\cal B}(f_{0} \rightarrow \pi^{+}\pi^{-})$, which are in agreement with the existing experimental values. Our results can help one to understand the internal structure of scalar mesons and the nature of tensor mesons and can be tested by future experiments.

\end{abstract}

\maketitle

%
%

\section{Introduction}\label{sec:intro}

Three-body hadronic $B$ meson decays can be important for testing the Standard Model (SM) by measuring the associated Cabibbo-Kobayashi Maskawa (CKM) matrix parameters such as the weak phases $\alpha$, $\beta$, and $\gamma$, understanding the potential mechanisms of hadronic weak decays and $CP$ violation, and discovering new physics\cite{Yan:2023yvx}. From the point of view of the spin-parity quantum numbers $J^{P}$, $J^{P}=2^{+}$, $1^{+}$, $0^{+}$, $0^{-}$ represents tensor mesons ($T$), vector mesons ($V$), scalar mesons ($S$) and pseudoscalar mesons ($P$), respectively\cite{Zou:2012td}. In this paper we study the tensors including isovector mesons $a_{2}(1320)$, isodoulet states $K^{*}_{2}(1430)$ and two isosinglet mesons $f_{2}(1270)$, $f^{'}_{2}(1525)$\cite{{Wang:2010ni},{ParticleDataGroup:2022pth}}. The factorizable amplitude with a tensor meson emitted is prohibited because a tensor meson cannot be created from the $ (V \pm A) $ or $ (S \pm P) $ currents\cite{Zou:2012td}. The isoscalar tensor states $f_{2}(1270)$ and $f^{'}_{2}(1525)$ have a mixing. From the experiment that $f_{2}(1270)$ decays dominantly into $\pi\pi$, and $KK$ is the dominant decay mode of $f^{'}_{2}(1525)$, we can know that the $f_{2}(1270)-f^{'}_{2}(1525)$ mixing angle $\phi$ is really small, $f_{2}(1270)$ is nearly $(u\bar{u}+d\bar{d})/\sqrt{2}$, and $f^{'}_{2}(1525)$ is mainly $s\overline{s}$\cite{Cheng:2011fk,ParticleDataGroup:2022pth,Li:2000zb}. The physical states $f_{2}(1270)$ and $f^{'}_{2}(1525)$ can be given by\cite{Zou:2012td}

\begin{equation}
\begin{split}
f_{2}(1270) = f^{n}_{2}\cos\phi + f^{s}_{2}\sin\phi ,\\
f^{'}_{2}(1525) = f^{n}_{2}\sin\phi-f^{s}_{2}\cos\phi ,
\end{split}
\end{equation}
we use the quark-flavor basis, with $f^{n}_{2}=\frac{1}{\sqrt{2}}(u\overline{u}+d\overline{d})$, $f^{s}_{2}=s\overline{s}$. Moreover, it is also theoretically found that the mixing angle $\phi=5.8^{\circ}$\cite{Cheng:2011fk}, $7.8^{\circ}$\cite{Li:2000zb}, $(9 \pm 1)^{\circ}$\cite{Li:2018lbd}. In two-body decays, there have been experimental and theoretical studies of $B \rightarrow PT, VT, ST, TT$\cite{Liu:2017cwl,Qin:2014xta,Li:2019jlp,Dai:2023knn}. In three-body decays, much work has been done on the tensor as a resonance state\cite{Chang:2024qxl,Fang:2023dcy,Liu:2021nun,Li:2020zng,Li:2019hnt}. For the first time, we treat the tensor as a nonresonant state, which is necessary to gain insight into the relevant decays and facilitates the exploration of the nature of the scalar resonances involved.

The identification of scalar resonances is a long-standing challenge. Scalar resonances are difficult to deal with because some of them have very broad decay widths, leading to strong overlap between the resonance and the background. In addition, the fundamental structure of scalar mesons is still quite controversial on the theoretical side. Currently, there are two main different scenarios, the so-called Scenario I (S-I) and Scenario II (S-II), to explain scalar mesons\cite{Cheng:2006yov,Ren:2023ebq}. In S-I, the mesons below or close to 1 GeV are considered as the lowest $q\bar{q}$ bound states, while mesons above 1 GeV are the first excited two-quark states. In contrast, in S-II, mesons near 1.5 GeV are regarded as ground two-quark states, while lighter mesons are identified with the dominant $qq\bar{q}\bar{q}$ state and possibly mixed with glueball states. Each scenario has its own physical meaning. In order to give quantitative predictions, we will only work in the S-I where $f_{0}(980)$ is regarded as the conventional two-quark $q\bar{q}$ state. The reason is that the $f_{0}(980)$ described by the four-quark state in S-II is too complex. It cannot be studied by factorization approaches\cite{Yan:2023yvx}. In addition, the presence of nonstrange and strange quark contents in $f_{0}(980)$ and $f_{0}(500)$ has been confirmed by the experimental measurements of the decays $D^{+}_{s} \rightarrow f_{0}\pi$, $\phi \rightarrow f_{0}\gamma$, $J/{\psi} \rightarrow f_{0}\omega$, $J/{\psi} \rightarrow f_{0}\phi$. $f_{0}(980)$ and $f_{0}(500)$ can be regarded as a mixture of $s\overline{s}$ and $(u\overline{u}+d\overline{d})/\sqrt{2}$\cite{E:2001th}. The mixing relation for the $f_{0}-\sigma$ system\cite{Gokalp:2004ny},

\begin{equation}
\begin{pmatrix}
\sigma \\
f_{0}
\end{pmatrix}
=
\begin{pmatrix}
\cos\theta & -\sin\theta \\
\sin\theta & \cos\theta
\end{pmatrix}
\begin{pmatrix}
f_{n} \\
f_{s}
\end{pmatrix}
\end{equation}
with $f_{n}=\frac{1}{\sqrt{2}}(u\overline{u}+d\overline{d}), f_{s}=s\overline{s}$. The mixing angle $\theta$ has not been determined experimentally or theoretically. The authors use the ratio between $J/{\psi} \rightarrow f_{0}\omega$ and $J/{\psi} \rightarrow f_{0}\phi$ to derive a mixing angle of $(34 \pm 6)^{\circ}$ and $(146 \pm 6)^{\circ}$\cite{Achasov:2002xd,Cheng:2002ai}. The WA102 experiment on $f_{0}(980)$ production in central pp collisions yields a result of $\theta = (42.3^{+8.3}_{-5.5})^{\circ}$ and $(158 \pm 2)^{\circ}$\cite{WA102:1999fqy}. Furthermore, a phenomenological analysis of the radiative decays $\phi \rightarrow f_{0}\gamma$ and $f_{0}(980) \rightarrow \gamma\gamma$ shows that the $\theta = (138 \pm 6)^{\circ}$ is preferable\cite{Cheng:2002ai}. Conservatively, we set $\theta$ to be a free parameter in this work.

It is well known that the multibody decay of heavy mesons involves more complex dynamics than the two-body decay and is difficult to describe in full phase space. It has been proposed that the factorization theorem for three-body $B$ decay is approximately valid when two particles move collinear and the bachelor particle recoils back, which provides a theoretical framework for the study of resonance contributions based on the quasi-two-body decay mechanism. Three-body $B$ decays have been extensively studied using various methods, such as those based on the symmetry principles\cite{He:2014xha}, the QCD factorization approach\cite{Krankl:2015fha,Cheng:2007si,Cheng:2016shb,Li:2014oca}, the perturbative QCD approach (pQCD)\cite{Wang:2016rlo,Li:2016tpn,Zou:2020ool,Li:2018psm} and so on. Previous studies have shown that the multibody $B$ meson decay branching ratio calculated by the pQCD approach based on the $k_{T}$ factorization fits well with experimental results, which further verifies the effectiveness of this method\cite{Yan:2024ymv,Li:2021qiw,Rui:2021kbn,Liang:2022mrz,Liang:2019eur}. In the pQCD approach, the three-body decay is simplified to the quasi-two-body decay and the pair of final state mesons is described by introducing the two-meson distribution amplitudes\cite{M:2000xyh}. The contribution from the direct evaluation of hard b-quark decay kernels containing two virtual gluons is generally power suppressed, and the dominant contribution comes from the region where the two energetic light mesons are almost collimating to each other with an invariant mass below $O(\overline{\Lambda}M_{B})$, ($\overline{\Lambda}=M_{B}-m_{b}$ is the difference in mass between the $B$ meson and $b$ quark). Then, the pQCD factorization formula for the three-body decay amplitude of the $B$ meson is written as\cite{C:2004cpa,C:2003vda}

\begin{equation}
{\cal A}={\cal H}\otimes \phi_{B} \otimes \phi_{h_{3}} \otimes \phi_{h_{1}h_{2}},
\end{equation}
where the hard decay kernel ${\cal H}$ can be calculated by using the perturbative theory. The nonperturbative inputs $\phi_{B}$, $\phi_{h_{1}h_{2}}$ and $\phi_{h_{3}}$ are the distribution amplitudes of the $B$ meson, $h_1h_2$ pair and $h_3$ meson respectively.

In the last decades, several $B$ decays with resonance states $a_{0}(980)$, $f_{0}(980)$ and $f_{0}(500)$ have been observed experimentally\cite{BABAR:2004ae,BABAR:2006siy,Belle:2006emv}, and the corresponding theoretical calculations have attracted increasing attention. For example, the LHCb Collaboration and the Belle Collaboration have observed $B^{0} \rightarrow J/{\psi}K^{+}K^{-}$ decay and $B^{\pm} \rightarrow K^{\pm}f_{0}(980) \rightarrow K^{\pm}{\pi}^{\mp}{\pi}^{\pm}$ decay with $a_{0}(980)$, $f_{0}(980)$ as resonance states, respectively\cite{Aaij:2013zpx,Belle:2002dka}. In Ref.\cite{R:2019ngy}, the branching ratios of the $B \rightarrow J/{\psi}(K\overline{K},\pi\eta)$ decays have been calculated with the resonances $a_{0}(980)$ and $a_{0}(1450)$, and the timelike form factors in the dimeson distribution amplitudes timelike form factors are parametrized by the Flatt$\acute{e}$ model and the Breit-Wigner formula, respectively. The authors of Ref.\cite{J:2022wa}, employing $a_{0}(980)$, $a_{0}(1450)$ and $a_{0}(1950)$ as resonances, studied the multiparticle configurations of isovector scalar mesons, in the $B \rightarrow a_{0}(\rightarrow K\overline{K},\pi\eta)h$ by considering the width effects. In Ref.\cite{Ma:2017idu},the Breit-Wigner formula and the Bugg$^{\prime}$s model for $f_{0}(500)$ resonance, and the Flatt$\acute{e}$ model for $f_{0}(980)$ resonance are adopted to study the quasi-two-body $B^{0}_{(s)} \rightarrow \eta_{c}(2S)\pi^{+}\pi^{-}$. In Ref.\cite{Zhang:2024ykn}, the author studied the quasi-two-body decays $B \rightarrow Pf_{0}(500) \rightarrow P\pi^{+}\pi^{-}$. Based on the decay $B^{0}_{s} \rightarrow SS(a_{0}(980),f_{0}(980),f_{0}(500))$\cite{Liang:2019eur}, the author predicted the branching ratios of the quasi-two-body decays $B^{0}_{s} \rightarrow S[\rightarrow P_{1}P_{2}]S$\cite{Yang:2022jog}. In Ref.\cite{W:2022uun} and Ref.\cite{L:2021vb}, the branching ratios of the $B \rightarrow K({\cal R} \rightarrow K^{+}K^{-})$ and $B_{(s)} \rightarrow V\pi\pi$ decays with $f_{0}(980)$ resonance have been studied by considering the mixing of $s\overline{s}$ and $(u\overline{u}+d\overline{d})/\sqrt{2}$. Combining the theoretical results\cite{Li:2019jlp} and the BABAR Collaboration measurements of $B^{0} \rightarrow K^{*}_{2}(1430) f_{0}(980)$\cite{BaBar:2011ryf}, we shall extend the study to $B^{0} \rightarrow TS[\rightarrow P_{1}P_{2}]$, where $T$ denotes the tensor mesons $a_{2}(1320)$, $K^{*}_{2}(1430)$, $f_{2}(1270)$ and $f^{'}_{2}(1525)$, and $P_{1}P_{2}=\pi\eta, \pi\pi, K\overline{K}$ is the final state meson pair.\footnote{$a_{0}$, $f_{0}$, and $\sigma$ refer to $a_{0}(980)$, $f_{0}(980)$, and $f_{0}(500)$ respectively in the following text. Moreover $a_{2}$, $f_{2}$, and $f^{'}_{2}$ refer to $a_{2}(1320)$, $f_{2}(1270)$, and $f^{'}_{2}(1525)$ respectively} For the mesons $f_{0}(980)$ and $f_{0}(500)$, $f_{2}(1270) $and $f^{'}_{2}(1525)$ we also adopt the corresponding mixing mechanism. The results presented in this paper can be validated in the LHCb and Belle II experiments in the near future.

This paper is organized as follows. In Sec \ref{sec:pert}, we introduce the theoretical framework. Section \ref{sec:numer} will give the numerical values and some discussion. Section \ref{sec:summary} contains our conclusions. The Appendix collects the explicit PQCD factorization formulas for all the decay amplitudes.

\begin{figure}[htbp]
	\centering
	\begin{tabular}{l}
		\includegraphics[width=0.8\textwidth]{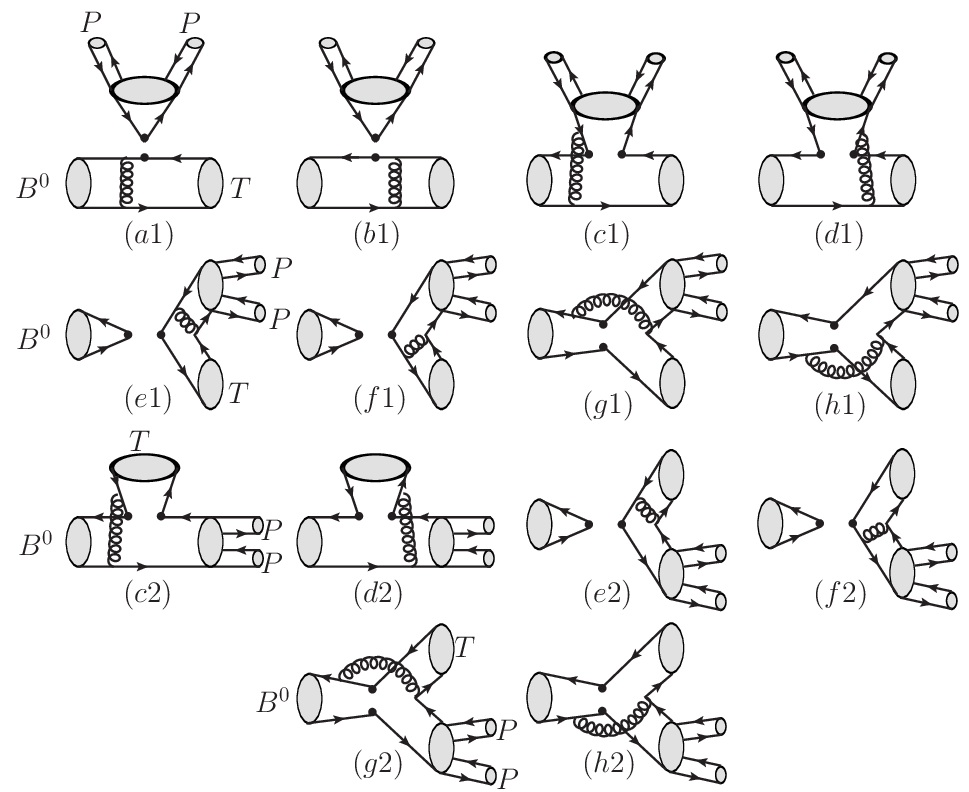}
	\end{tabular}
	\caption{The Feynman diagrams for the $B^{0} \rightarrow TS[\rightarrow{P_{1}P_{2}}]$ decays in pQCD. The symbol Black dot stands for the weak vertex, $T$ means the tensor mesons, and $P_{1}P_{2}$ denotes the final state meson pair; the corresponding relationship are $a_{0}(\pi\eta, K\overline{K})$, $f_{0}(\pi\pi, K\overline{K})$ and $\sigma(\pi\pi)$.}
	\label{fig:figure1}
\end{figure}

%
%
\section{The Theoretical Framework and Helicity Amplitudes}\label{sec:pert}

\subsection{The wave functions}

The relevant weak effective Hamiltonian of the quasi-two-body $B^{0} \rightarrow TS[ \rightarrow P_{1}P_{2}]$ decays can be written as\cite{Buchalla:1995vs}

\begin{equation}
{\cal H}_{eff}=\frac{G_{F}}{\sqrt2}\big\{V^{*}_{ub}V_{uq}[C_{1}(\mu)O_{1}(\mu)+C_{2}(\mu)O_{2}(\mu)]-V^{*}_{tb}V_{tq}[\sum^{10}_{i=3}C_{i}(\mu)O_{i}(\mu)]\big\},
\end{equation}
where the $V^{*}_{ub}V_{uq}$ and CKM matrix elements with $q = d, s$ quark, the Fermi constant $G_{F}$=$1.66378\times10^{-5}{\rm GeV^{-2}}$, and $C_{i}$ is the corresponding Wilson coefficient. The local four-quark operator $O_{i}(\mu)$ can be divided into the following three categories and the relevant Feynman diagrams illustrated in Fig.~\ref{fig:figure1}.

\begin{itemize}
\item[$\bullet$] Current-current (tree) operators:
\begin{equation}
\begin{split}\\
O_{1}=(\overline{b}_{\alpha}u_{\beta})_{V-A}(\overline{u}_{\beta}q_{\alpha})_{V-A},
O_{2}=(\overline{b}_{\alpha}u_{\alpha})_{V-A}(\overline{u}_{\beta}q_{\beta})_{V-A},
\end{split}
\end{equation}

\item[$\bullet$] QCD penguin operators:
\begin{equation}
\begin{split}\\
O_{3}&=(\overline{b}_{\alpha}q_{\alpha})_{V-A}\sum_{q^{\prime}}(\bar{q}^{\prime}_{\beta}q^{\prime}_{\beta})_{V-A},
O_{4}=(\overline{b}_{\alpha}q_{\beta})_{V-A}\sum_{q^{\prime}}(\bar{q}^{\prime}_{\beta}q^{\prime}_{\alpha})_{V-A},\\
O_{5}&=(\overline{b}_{\alpha}q_{\alpha})_{V-A}\sum_{q^{\prime}}(\overline{q}^{\prime}_{\beta}q^{\prime}_{\beta})_{V+A},
O_{6}=(\overline{b}_{\alpha}q_{\beta})_{V-A}\sum_{q^{\prime}}(\overline{q}^{\prime}_{\beta}q^{\prime}_{\alpha})_{V+A},\\
\end{split}
\end{equation}

\item[$\bullet$] Electroweak penguin operators:
\begin{equation}
\begin{split}\\
O_{7}&=\frac{3}{2}(\overline{b}_{\alpha}q_{\alpha})_{V-A}\sum_{q^{\prime}}e_{q^{\prime}}(\overline{q}^{\prime}_{\beta}q^{\prime}_{\beta})_{V+A},
O_{8}=\frac{3}{2}(\overline{b}_{\alpha}q_{\beta})_{V-A}\sum_{q^{\prime}}e_{q^{\prime}}(\overline{q}^{\prime}_{\beta}q^{\prime}_{\alpha})_{V+A},\\
O_{9}&=\frac{3}{2}(\overline{b}_{\alpha}q_{\alpha})_{V-A}\sum_{q^{\prime}}e_{q^{\prime}}(\overline{q}^{\prime}_{\beta}q^{\prime}_{\beta})_{V-A},
O_{10}=\frac{3}{2}(\overline{b}_{\alpha}q_{\beta})_{V-A}\sum_{q^{\prime}}e_{q^{\prime}}(\overline{q}^{\prime}_{\beta}q^{\prime}_{\alpha})_{V-A},
\end{split}
\end{equation}
\end{itemize}
where the subscripts $\alpha$ and $\beta$ are the color indices and $q^{\prime}$ are the active quarks at the scale $m_b$, i.e, $q^{\prime}= (u, d, s, c, b)$. The left-handed current is defined as $(\overline{b}_{\alpha}q_{\alpha})_{V-A}=\overline{b}_{\alpha}\gamma_{\mu}(1-\gamma_{5})q_{\alpha}$, and the right-handed current is defined as $(\overline{q}^{\prime}_{\beta}q^{\prime}_{\alpha})_{V+A}=\overline{q}^{\prime}_{\beta}\gamma_{\mu}(1+\gamma_{5})q^{\prime}_{\alpha}$.

In the light cone coordinates, we let the $B^0$ meson stay at rest, and choose the $P_{1}P_{2}$ meson pair, and the final-state $T$ move along the direction of $n=(1,0,0_{T})$ and $v=(0,1,0_{T})$, respectively. Thus, the momentum $p_{B}$ of the $B^0$ meson, the total momentum $p=p_{1}+p_{2}$ of the $P_{1}P_{2}$ meson pair, and the momentum $p_{3}$ of the final state $T$ are respectively

\begin{equation}
\begin{split}
&p_{B}=\frac{M_{B}}{\sqrt{2}}(1,1,0_{T}),                 \\
&p=\frac{M_{B}}{\sqrt{2}}(1-r^2,\eta,0_{T}),              \\
&p_{3}=\frac{M_{B}}{\sqrt{2}}(r^2,1-\eta,0_{T}),
\end{split}
\end{equation}
where $M_{B}$ represents the mass of $B^{0}$, $r=\frac{m_{T}}{M_{B}}$ is the mass ratio, and $m_{T}$ refers to the mass of the final-state $T$. We think the variable $\eta=\omega^2/({M_{B}^2}-{m_{T}^2})$, and $\omega$ is the invariant mass of the $P_{1}P_{2}$ meson pair, which satisfies the relation $\omega^2=p^2$. Meanwhile, we define $\zeta=p^+_{1}/p^+$ as one of the $P_{1}P_{2}$ meson pair's momentum fractions. Accordingly, the kinematic variables of other components in the meson pair can be expressed as

\begin{equation}
\begin{split}
& p^-_{1}=\frac{M_{B}}{\sqrt{2}}(1-\zeta)\eta ,  \\
&p^+_{2}=\frac{M_{B}}{\sqrt{2}}(1-\zeta)(1-r^2),\\
&p^-_{2}=\frac{M_{B}}{\sqrt{2}}\zeta\eta .
\end{split}
\end{equation}

We adopt $x_{B}$, $z$, $x_{3}$ to indicate the momentum fraction of the light quark in each meson with the range from zero to unity. So the light quark's momentum of the $B^0(k_{B})$, $P_{1}P_{2}(k)$ and $T(k_{3})$ are defined as

\begin{equation}
\begin{split}
&k_{B}=(0,\frac{M_{B}}{\sqrt{2}}{x_{B}},k_{BT}),\\
&k=(\frac{M_{B}}{\sqrt{2}}z(1-r^2),0,k_{T}),\\
&k_{3}=(\frac{M_{B}}{\sqrt{2}}r^2x_{3},\frac{M_{B}}{\sqrt{2}}(1-\eta)x_{3},k_{3T}).
\end{split}
\end{equation}

For the $B^{0}$ meson, the wave function can be expressed as\cite{C:2001oqa,Keum:2000wi,HnLijhep2013}
\begin{equation}
\Phi_B=\frac{i}{\sqrt{2N_{c}}}(\not {p}_{B}+M_{B}){\gamma_{5}}{\phi_{B}({x_{B},b_{B}})},
\end{equation}
where $N_{c}=3$ is the color factor. The light cone distribution amplitude(LCDA) ${\phi_{B}({x_{B},b_{B}})}$ is explicitly expressed as
\begin{equation}
\phi_{B}(x_{B},b_{B})=\emph{N}_{B}{{x_{B}}^2}(1-{x_{B}})^2\exp[-\frac{M^2_{B}{{x_{B}}^2}}{2\omega^2_{b}}-\frac{1}{2}(\omega_{b}{b_{B}})^2],
\end{equation}
where the $\emph{N}_{B}$ is the normalization constant, which can be determined by the normalization condition
\begin{equation}
\int^{1}_{0}\phi_{B}(x_{B},b_{B}=0)\emph{dx}=\emph{f}_{B}/({2}{\sqrt{{2}{\emph{N}}_{c}}}),
\end{equation}
with the decay constant $f_{B} = (0.19 \pm 0.02)$GeV. The corresponding shape parameter $\omega_{b}$ in the LCDA of the $B^{0}$ meson is usually taken as the value $(0.40 \pm 0.04)$GeV.

In the quark model, the tensor meson with $J^{PC}=2^{++}$ has angular momentum $L = 1$ and spin $S = 1$. The polarization of $\lambda = \pm2$ vanishes in the three-body decays $B^{0} \rightarrow TS[\rightarrow P_{1}P_{2}]$ because of angular momentum conservation. In this case, the wave function of the tensor meson is very similar to the vector meson and can be defined as\cite{Cheng:2010hn,Cheng:2010yd}

\begin{equation}
\begin{split}\label{equ:d}
\Phi_T&=\frac{1}{\sqrt{2N_{c}} }[m_{T}\not {\epsilon}^{*}_{\bullet L}\phi_{T}(x)+\not {\epsilon}^{*}_{\bullet L}\not{P}\phi^{t}_{T}(x)+m^{2}_{T}\frac{\epsilon_{\bullet}\cdot {v}}{P\cdot{v}}\phi^{s}_{T}(x)]\\
\Phi^{\perp}_T&=\frac{1}{\sqrt{2N_{c}} }[m_{T}\not {\epsilon}^{*}_{\bullet \perp}\phi^{v}_{T}(x)+\not {\epsilon}^{*}_{\bullet \perp}\not{P}\phi^{T}_{T}(x)+m_{T}i\epsilon_{\mu\nu\rho\sigma}\gamma_{5}\gamma^{\mu}\epsilon^{*\nu}_{\bullet \perp}n^{\rho}v^{\sigma}\phi^{a}_{T}(x)]\\
\end{split}
\end{equation}
with $\epsilon^{0123}=1 $. The reduced polarization vector $\epsilon_{\bullet\mu}=\frac{\epsilon_{\mu \nu}v^{\nu}}{P\cdot{v}}$, in which
$\epsilon_{\mu \nu}$ is the polarization tensor of the tensor meson.
The expressions of twist-2 and twist-3 LCDAs are given as
\begin{equation}
\begin{split}
\phi_{T}(x)&=\frac{f_{T}}{2\sqrt{2N_{c}}}\phi_{\parallel}(x), \phi^{t}_{T}(x)=\frac{f^{\perp}_{T}}{2\sqrt{2N_{c}}}h^{t}_{\parallel}(x),\\
\phi^{s}_{T}(x)&=\frac{f^{\perp}_{T}}{4\sqrt{2N_{c}}}\frac{d}{dx}h^{s}_{\parallel}(x), \phi^{T}_{T}(x)=\frac{f^{\perp}_{T}}{2\sqrt{2N_{c}}}\phi_{\perp}(x),\\
\phi^{v}_{T}(x)&=\frac{f_{T}}{2\sqrt{2N_{c}}}g^{v}_{\perp}(x), \phi^{a}_{T}(x)=\frac{f_{T}}{8\sqrt{2N_{c}}}\frac{d}{dx}g^{a}_{\perp}(x).\\
\end{split}
\end{equation}
with the detailed

\begin{equation}
\begin{split}
\phi_{\parallel,\perp}(x)&=30x(1-x)(2x-1), g^{v}_{\perp}(x) = 5(2x-1)^{3},\\
 h^{t}_{\parallel}(x)& =\frac{15}{2}(2x-1)(1-6x+6x^{2}) ,\\
h^{s}_{\parallel}(x) &= 15x(1-x)(2x-1), g^{a}_{\perp}(x) = 20x(1-x)(2x-1).\\
\end{split}
\end{equation}

In this work, the S-wave two-meson DA is written in the form

\begin{equation}
\Phi^{S}_{P_{1}P_{2}}=\frac{1}{\sqrt{2N_{c}}}[\not{p}\phi_{S}(z,\zeta,\omega^{2})+\omega\phi^{s}_{S}(z,\zeta,\omega^{2})+\omega(\not{n}\not{v}-1)\phi^{t}_{S}(z,\zeta,\omega^{2})],
\end{equation}
and the asymptotic forms of the individual twist-2 and twist-3 DAs $\phi_{S}$ and  $\phi^{s,t}_{S}$ are parametrized as
\begin{equation}
\begin{split}
&\phi_{S}(z,\zeta,\omega^{2})=\frac{9F_{S}(\omega)}{\sqrt{2N_{c}}}a_{2}z(1-z)(1-2z),\\
&\phi^{s}_{S}(z,\zeta,\omega^{2})=\frac{F_{S}(\omega)}{2\sqrt{2N_{c}}},\\
&\phi^{t}_{S}(z,\zeta,\omega^{2})=\frac{F_{S}(\omega)}{2\sqrt{2N_{c}}}(1-2z).
\end{split}
\end{equation}
with the Gegenbauer coefficient $a_{2}=0.3\pm0.1$ for $a_{0}$ and $a_{2}=0.3\pm0.2$ for $f_{0}$\cite{Aaij:2013zpx,L:2021vb,Ye:2019qu}. $F_{S}(\omega)$ is the timelike form factor, which can factorize the strong interactions between the resonance and the final-state meson pair, as well as the elastic rescattering processes of the final-state meson pair.

For the scalar resonances $a_{0}$ and $f_{0}$, we adopt the Flatt$\acute{e}$ parametrization where the resulting line shape is above and below the threshold of the intermediate particle. The pole masses of the $a_{0}$ and $f_{0}$ resonances are both close to the $K\overline{K}$ threshold, and the main decay channels are $a_{0}\rightarrow\pi\eta,K\overline{K}$ and $f_{0}\rightarrow \pi\pi, K\overline{K}$. Since theFlatt$\acute{e}$ parametrization shows a scaling invariance and does not allow for an extraction of individual partial decay widths when the coupling of a resonance to the channel opening nearby is very strong, we employ the modified Flatt$\acute{e}$ model

\begin{equation}
F^{f_{0}}_{S}(\omega)=\frac{m^{2}_{f_{0}}}{m^{2}_{f_{0}}-\omega^{2}-im_{f_{0}}(g_{\pi\pi}\rho_{\pi\pi}+g_{KK}\rho_{KK}F^{2}_{KK})},
\end{equation}
for the $f_{0}$ resonance\cite{Zou:2020ool,L:2021vb} and

\begin{equation}
F^{a_{0}}_{S}(\omega)=\frac{C_{a_{0}}m^{2}_{a_{0}}}{m^{2}_{a_{0}}-\omega^{2}-i(g^{2}_{\pi\eta}\rho_{\pi\eta}+g^{2}_{KK}\rho_{KK})},
\end{equation}
for the $a_{0}$ resonance\cite{R:2019ngy}. The timelike form factor of the resonant state $a_{0}$ is parametrized by the complex amplitude $C_{a_{0}}=|C_{a_{0}}|e^{i\phi_{a_{0}}}$, where the value depends on whether the final state mesons are $\pi\eta$ or $K\overline{K}$. For the $a_{0} \rightarrow K\overline{K}$ channel, the magnitude of the complex amplitude $|C^{KK}_{a_{0}}|=1.07$ and the phase angle $\phi_{a_{0}}=82^{\circ}$. When the channel is $a_{0} \rightarrow \pi\eta$, the phase angle remains the same as in the $K\overline{K}$ system, and the amplitude magnitude satisfies $C^{\pi\eta}_{a_{0}}/C^{KK}_{a_{0}}=g_{a_{0}\pi\eta}/g_{a_{0}KK}$. The definition of the strong coupling constants $g_{a_{0}KK}(g_{a_{0}\pi\eta})$ can be found in the literature\cite{J:2022wa,Wang:2020saq}. We determined the strong coupling constants $g_{a_{0}KK}$ and $g_{a_{0}\pi\eta}$ by using the the relation $g_{a_{0}KK}(g_{a_{0}\pi\eta})/(4\sqrt{\pi})=g_{KK}(g_{\pi\eta})$ and the coupling constants $g_{\pi\eta}=0.324$ GeV, $g^{2}_{KK}/g^{2}_{\pi\eta}=1.03$ from the Crystal Barrel experiment. Meanwhile, we employ the coupling constants $g_{\pi\pi}=0.165\pm0.018$ GeV and $g_{KK}/g_{\pi\pi}=4.21\pm0.33$ for $f_{0}$\cite{Back:2018fqe,L:2021vb}, and introduce the factor $F_{KK}=e^{{-\alpha}q^{2}}$ into the timelike form factor $F^{f_{0}}_{S}(\omega)$ to suppress the $K\overline{K}$ contribution with the parameter $\alpha=2.0\pm1.0$ GeV$^{-2}$. In addition, the $\rho$ factors are chosen as:

\begin{equation}
\rho_{\pi\eta}=\sqrt{[1-(\frac{m_{\eta}-m_{\pi}}{\omega})^{2}][1-(\frac{m_{\eta}+m_{\pi}}{\omega})^{2}]},
\end{equation}

\begin{equation}
\rho_{\pi\pi}=\frac{2}{3}\sqrt{1-\frac{4m^{2}_{\pi^{\pm}}}{\omega^{2}}}+\frac{1}{3}\sqrt{1-\frac{4m^{2}_{\pi^{0}}}{\omega^{2}}},
\end{equation}

\begin{equation}
\rho_{KK}=\frac{1}{2}\sqrt{1-\frac{4m^{2}_{K^{\pm}}}{\omega^{2}}}+\frac{1}{2}\sqrt{1-\frac{4m^{2}_{K^{0}}}{\omega^{2}}}.
\end{equation}
Following the analysis of the LHCb Collaboration\cite{R:2013az,R:2014uea,LHCb:2015klp}, and motivated by the studies in Refs\cite{Liang:2022mrz,Ma:2017idu,Yang:2022jog,Li:2015tja,Wang:2015uea}, the shape of the $\sigma$ resonance can be well described by the Breit-Wigner model:

\begin{equation}
F^{\sigma}_{S}(\omega)=\frac{C_{\sigma}m^{2}_{\sigma}}{m^{2}_{\sigma}-\omega^{2}-im_{\sigma}\Gamma(\omega)},
\end{equation}
with the factor $C_{\sigma}=3.50$. The energy-dependent width $\Gamma(\omega)$ in the case of a scalar resonance decaying into pion pair can be parametrized as

\begin{equation}
\Gamma(\omega)=\Gamma_{0}\frac{m_{\sigma}}{\omega}(\frac{\omega^{2}-4m^{2}_{\pi}}{m^{2}_{\sigma}-4m^{2}_{\pi}})^{\frac{1}{2}},
\end{equation}
where $\Gamma_{0}=0.40$ GeV is the width of the resonance.

\subsection{Helicity amplitudes}

According to the typical Feynman diagrams as shown in Fig.~\ref{fig:figure1}, the total decay amplitude for each considered decay mode in this work is given as follows:

\begin{equation}
\begin{split}
{\cal A}(B^{0} \rightarrow K^{*}_{2} f_{s}[ \rightarrow \pi^{+}\pi^{-}, K^{+}K^{-}])=
&-\frac{G_{F}}{\sqrt{2}}V^{*}_{tb}V_{ts}[(\frac{1}{3}C_{5}-\frac{1}{6}C_{7}+C_{6}-\frac{1}{2}C_{8})(F^{SP}_{S}+A^{SP}_{S})\\
&+(\frac{1}{3}C_{3}-\frac{1}{6}C_{9}+C_{4}-\frac{1}{2}C_{10})A^{LL}_{S}+(C_{6}-\frac{1}{2}C_{8})M^{SP}_{S}\\
&+(C_{3}-\frac{1}{2}C_{9}+C_{4}-\frac{1}{2}C_{10})M^{LL}_{S}+(C_{3}-\frac{1}{2}C_{9})W^{LL}_{S}\\
&+(C_{5}-\frac{1}{2}C_{7})(M^{LR}_{S}+W^{LR}_{S})],
\end{split}
\end{equation}

\begin{equation}
\begin{split}
{\cal A}(B^{0} \rightarrow K^{*}_{2} f_{n}[ \rightarrow \pi^{+}\pi^{-}, K^{+}K^{-}])=
&\frac{G_{F}}{2}\{V^{*}_{ub}V_{us}C_{2}M^{LL}_{S}-V^{*}_{tb}V_{ts}[(C_{4}+\frac{1}{3}C_{3}-\frac{1}{2}C_{10}-\frac{1}{6}C_{9})A^{LL}_{T}\\
&+(C_{6}+\frac{1}{3}C_{5}-\frac{1}{2}C_{8}-\frac{1}{6}C_{7})A^{SP}_{T}+(C_{3}-\frac{1}{2}C_{9})(M^{LL}_{T}+W^{LL}_{T})\\
&+(C_{5}-\frac{1}{2}C_{7})(M^{LR}_{T}+W^{LR}_{T})+(2C_{6}+\frac{1}{2}C_{8})M^{SP}_{S}\\
&+(2C_{4}+\frac{1}{2}C_{10})M^{LL}_{S}]\},
\end{split}
\end{equation}

\begin{equation}
\begin{split}
{\cal A}(B^{0} \rightarrow K^{*}_{2}a^{0}_{0}[ \rightarrow \pi^{0}\eta, K^{+}K^{-}])=
&\frac{G_{F}}{2}\{V^{*}_{ub}V_{us}C_{2}M^{LL}_{S}-V^{*}_{tb}V_{ts}[(-C_{4}-\frac{1}{3}C_{3}+\frac{1}{2}C_{10}+\frac{1}{6}C_{9})A^{LL}_{T}\\
&+(-C_{6}-\frac{1}{3}C_{5}+\frac{1}{2}C_{8}+\frac{1}{6}C_{7})A^{SP}_{T}+(-C_{3}+\frac{1}{2}C_{9})(M^{LL}_{T}+W^{LL}_{T})\\
&+(-C_{5}+\frac{1}{2}C_{7})(M^{LR}_{T}+W^{LR}_{T})+\frac{3}{2}C_{8}M^{SP}_{S}+\frac{3}{2}C_{10}M^{LL}_{S}]\},
\end{split}
\end{equation}

\begin{equation}
\begin{split}
{\cal A}(B^{0} \rightarrow K^{*+}_{2}a^{-}_{0}[ \rightarrow \pi^{-}\eta, K^{-}\overline{K}^{0}])=
&\frac{G_{F}}{\sqrt{2}}\{V^{*}_{ub}V_{us}C_{1}M^{LL}_{T}-V^{*}_{tb}V_{ts}[(C_{4}+\frac{1}{3}C_{3}-\frac{1}{2}C_{10}-\frac{1}{6}C_{9})A^{LL}_{T}\\
&+(C_{6}+\frac{1}{3}C_{5}-\frac{1}{2}C_{8}-\frac{1}{6}C_{7})A^{SP}_{T}+(C_{3}-\frac{1}{2}C_{9})W^{LL}_{T}\\
&+(C_{5}-\frac{1}{2}C_{7})W^{LR}_{T}+(C_{3}+C_{9})M^{LL}_{T}+(C_{5}+C_{7})M^{LR}_{T}]\},
\end{split}
\end{equation}

\begin{equation}
\begin{split}
{\cal A}(B^{0} \rightarrow a^{0}_{2} f_{s}[ \rightarrow \pi^{+}\pi^{-}, K^{+}K^{-}])=
&-\frac{G_{F}}{2}V^{*}_{tb}V_{td}[(\frac{1}{2}C_{10}-C_{4})M^{LL}_{S}+(\frac{1}{2}C_{8}-C_{6})M^{SP}_{S}],\\
\end{split}
\end{equation}

\begin{equation}
\begin{split}
{\cal A}(B^{0} \rightarrow a^{0}_{2} f_{n}[ \rightarrow \pi^{+}\pi^{-}, K^{+}K^{-}])=
&\frac{G_{F}}{2\sqrt{2}}\{V^{*}_{ub}V_{ud}[(C_{1}+\frac{1}{3}C_{2})(A^{LL}_{S}+A^{LL}_{T})\\
&+C_{2}(M^{LL}_{T}-M^{LL}_{S}+W^{LL}_{T}+W^{LL}_{S})]\\
&-V^{*}_{tb}V_{td}[(-\frac{1}{3}C_{3}+\frac{3}{2}C_{7}+\frac{5}{3}C_{9}-C_{4}+\frac{1}{2}C_{8}+C_{10})(A^{LL}_{S}+A^{LL}_{T})\\
&+(\frac{1}{6}C_{7}-\frac{1}{3}C_{5}-C_{6}+\frac{1}{2}C_{8})(A^{SP}_{S}+A^{SP}_{T}+F^{SP}_{S})\\
&+(\frac{1}{2}C_{9}+\frac{3}{2}C_{10}-C_{3})(M^{LL}_{T}+W^{LL}_{T}+W^{LL}_{S})-(2C_{6}+\frac{1}{2}C_{8})M^{SP}_{S}\\
&+(\frac{1}{2}C_{9}-\frac{1}{2}C_{10}-C_{3}-2C_{4})M^{LL}_{S}+\frac{3}{2}C_{8}(M^{SP}_{T}+W^{SP}_{T}+W^{SP}_{S})\\
&+(\frac{1}{2}C_{7}-C_{5})(M^{LR}_{T}+M^{LR}_{S}+W^{LR}_{T}+W^{LR}_{S})]\},
\end{split}
\end{equation}

\begin{equation}
\begin{split}
{\cal A}(B^{0} \rightarrow a^{0}_{2}a^{0}_{0}[ \rightarrow \pi^{0}\eta, K^{+}K^{-}])=
&\frac{G_{F}}{2\sqrt{2}}\{V^{*}_{ub}V_{ud}[(C_{1}+\frac{1}{3}C_{2})(A^{LL}_{S}+A^{LL}_{T})+C_{2}(W^{LL}_{T}+W^{LL}_{S}-M^{LL}_{T}-M^{LL}_{S})]\\
&-V^{*}_{tb}V_{td}[-\frac{3}{2}C_{8}(M^{SP}_{T}+M^{SP}_{S})+(C_{3}-\frac{1}{2}C_{9}-\frac{3}{2}C_{10})(M^{LL}_{T}+M^{LL}_{S})]\\
&+(\frac{7}{3}C_{3}+\frac{1}{3}C_{9}+\frac{5}{3}C_{4}+\frac{2}{3}C_{6}+\frac{1}{6}C_{8}-\frac{1}{3}C_{10}+\frac{1}{2}C_{7}+2C_{5})(A^{LL}_{S}+A^{LL}_{T})\\
&+(C_{3}-\frac{1}{2}C_{9}+\frac{1}{2}C_{10}+2C_{4})(W^{LL}_{S}+W^{LL}_{T})+(\frac{1}{2}C_{8}+2C_{6})(W^{SP}_{S}+W^{SP}_{T})\\
&+(\frac{1}{3}C_{5}-\frac{1}{6}C_{7}-\frac{1}{2}C_{8}+C_{6})(A^{SP}_{S}+A^{SP}_{T}+F^{SP}_{S})\\
&+(C_{5}-\frac{1}{2}C_{7})(W^{LR}_{T}+W^{LR}_{S}+M^{LR}_{T}+M^{LR}_{S})\},
\end{split}
\end{equation}

\begin{equation}
\begin{split}
{\cal A}(B^{0} \rightarrow a^{+}_{2}a^{-}_{0}[ \rightarrow \pi^{-}\eta, K^{-}\overline{K}^{0}])=
&\frac{G_{F}}{\sqrt{2}}\{V^{*}_{ub}V_{ud}[(C_{1}+\frac{1}{3}C_{2})A^{LL}_{S}+C_{2}W^{LL}_{S}+C_{1}M^{LL}_{T}]\\
&-V^{*}_{tb}V_{td}[(C_{4}+C_{10})W^{LL}_{S}+(C_{6}+C_{8})W^{SP}_{S}\\
&+(C_{3}+\frac{1}{3}C_{4}+C_{5}+\frac{1}{3}C_{6}+C_{7}+\frac{1}{3}C_{8}+C_{9}+\frac{1}{3}C_{10})A^{LL}_{S}\\
&+(C_{3}+C_{9})M^{LL}_{T}+(C_{5}+C_{7})M^{LR}_{T}+(C_{5}-\frac{1}{2}C_{7})W^{LR}_{T}\\
&+(\frac{4}{3}C_{3}+\frac{4}{3}C_{4}-\frac{2}{3}C_{9}-\frac{2}{3}C_{10}+C_{5}+\frac{1}{3}C_{6}-\frac{1}{2}C_{7}-\frac{1}{6}C_{8})A^{LL}_{T}\\
&+(\frac{1}{3}C_{5}+C_{6}-\frac{1}{6}C_{7}-\frac{1}{2}C_{8})A^{SP}_{T}+(C_{6}-\frac{1}{2}C_{8})W^{SP}_{T}\\
&+(C_{3}+C_{4}-\frac{1}{2}C_{9}-\frac{1}{2}C_{10})W^{LL}_{T}]\},
\end{split}
\end{equation}

\begin{equation}
\begin{split}
{\cal A}(B^{0} \rightarrow a^{-}_{2}a^{+}_{0}[ \rightarrow \pi^{+}\eta, K^{+}\overline{K}^{0}])=
&\frac{G_{F}}{\sqrt{2}}\{V^{*}_{ub}V_{ud}[(C_{1}+\frac{1}{3}C_{2})A^{LL}_{T}+C_{2}W^{LL}_{T}+C_{1}M^{LL}_{S}]\\
&-V^{*}_{tb}V_{td}[(C_{4}+C_{10})W^{LL}_{T}+(C_{6}+C_{8})W^{SP}_{T}\\
&+(C_{3}+\frac{1}{3}C_{4}+C_{5}+\frac{1}{3}C_{6}+C_{7}+\frac{1}{3}C_{8}+C_{9}+\frac{1}{3}C_{10})A^{LL}_{T}\\
&+(C_{3}+C_{9})M^{LL}_{S}+(C_{5}+C_{7})M^{LR}_{S}+(C_{5}-\frac{1}{2}C_{7})W^{LR}_{S}\\
&+(\frac{4}{3}C_{3}+\frac{4}{3}C_{4}-\frac{2}{3}C_{9}-\frac{2}{3}C_{10}+C_{5}+\frac{1}{3}C_{6}-\frac{1}{2}C_{7}-\frac{1}{6}C_{8})A^{LL}_{S}\\
&+(\frac{1}{3}C_{5}+C_{6}-\frac{1}{6}C_{7}-\frac{1}{2}C_{8})A^{SP}_{S}+(C_{6}-\frac{1}{2}C_{8})W^{SP}_{S}\\
&+(C_{3}+C_{4}-\frac{1}{2}C_{9}-\frac{1}{2}C_{10})W^{LL}_{S}+(C_{6}+C_{8}+\frac{1}{3}C_{5}+\frac{1}{3}C_{7})F^{SP}_{S}]\},
\end{split}
\end{equation}

\begin{equation}
\begin{split}
{\cal A}(B^{0} \rightarrow f^{s}_{2} f_{s}[ \rightarrow \pi^{+}\pi^{-}, K^{+}K^{-}])=
&-\frac{G_{F}}{\sqrt{2}}V^{*}_{tb}V_{td}[(C_{3}+\frac{1}{3}C_{4}+C_{5}+\frac{1}{3}C_{6}-\frac{1}{2}C_{7}-\frac{1}{6}C_{8}-\frac{1}{2}C_{9}-\frac{1}{6}C_{10})\\
&(A^{LL}_{T}+A^{LL}_{S})+(C_{4}-\frac{1}{2}C_{10})(W^{LL}_{S}+W^{LL}_{T})\\
&+(C_{6}-\frac{1}{2}C_{8})(W^{SP}_{S}+W^{SP}_{T})],\\
\end{split}
\end{equation}

\begin{equation}
\begin{split}
{\cal A}(B^{0} \rightarrow f^{s}_{2} f_{n}[ \rightarrow \pi^{+}\pi^{-}, K^{+}K^{-}])=
&-\frac{G_{F}}{2}V^{*}_{tb}V_{td}[(C_{4}-\frac{1}{2}C_{10})M^{LL}_{T}+(C_{6}-\frac{1}{2}C_{8})M^{SP}_{T}],\\
\end{split}
\end{equation}

\begin{equation}
\begin{split}
{\cal A}(B^{0} \rightarrow f^{n}_{2} f_{s}[ \rightarrow \pi^{+}\pi^{-}, K^{+}K^{-}])=
&-\frac{G_{F}}{2}V^{*}_{tb}V_{td}[(C_{4}-\frac{1}{2}C_{10})M^{LL}_{S}+(C_{6}-\frac{1}{2}C_{8})M^{SP}_{T}],\\
\end{split}
\end{equation}

\begin{equation}
\begin{split}
{\cal A}(B^{0} \rightarrow f^{n}_{2} f_{n}[ \rightarrow \pi^{+}\pi^{-}, K^{+}K^{-}])=
&\frac{G_{F}}{2\sqrt{2}}\{V^{*}_{ub}V_{ud}[(C_{1}+\frac{1}{3}C_{2})(A^{LL}_{S}+A^{LL}_{T})+C_{2}(M^{LL}_{T}+M^{LL}_{S}+W^{LL}_{T}+W^{LL}_{S})]\\
&-V^{*}_{tb}V_{td}[(\frac{7}{3}C_{3}+2C_{5}+\frac{1}{2}C_{7}+\frac{1}{3}C_{9}+\frac{5}{3}C_{4}-\frac{1}{3}C_{10}+\frac{2}{3}C_{6}+\frac{1}{6}C_{8})(A^{LL}_{S}+A^{LL}_{T})\\
&+(\frac{1}{3}C_{5}+C_{6}-\frac{1}{6}C_{7}-\frac{1}{2}C_{8})(A^{SP}_{S}+A^{SP}_{T}+F^{SP}_{S})\\
&+(C_{3}+2C_{4}-\frac{1}{2}C_{9}+\frac{1}{2}C_{10})(M^{LL}_{T}+M^{LL}_{S}+W^{LL}_{T}+W^{LL}_{S})\\
&+(2C_{6}+\frac{1}{2}C_{8})(M^{SP}_{S}+M^{SP}_{T}+W^{SP}_{T}+W^{SP}_{S})\\
&+(C_{5}-\frac{1}{2}C_{7})(M^{LR}_{T}+M^{LR}_{S}+W^{LR}_{T}+W^{LR}_{S})]\},
\end{split}
\end{equation}

\begin{equation}
\begin{split}
{\cal A}(B^{0} \rightarrow f^{n}_{2}a^{0}_{0}[ \rightarrow \pi^{0}\eta, K^{+}K^{-}])=
&\frac{G_{F}}{2\sqrt{2}}\{V^{*}_{ub}V_{ud}[(C_{1}+\frac{1}{3}C_{2})(A^{LL}_{S}+A^{LL}_{T})+C_{2}(W^{LL}_{T}+W^{LL}_{S}-M^{LL}_{T}+M^{LL}_{S})]\\
&-V^{*}_{tb}V_{td}[\frac{3}{2}C_{8}(M^{SP}_{S}+W^{SP}_{T}+W^{SP}_{S})+(\frac{1}{2}C_{9}+\frac{3}{2}C_{10}-C_{3})(M^{LL}_{S}+W^{LL}_{S}+W^{LL}_{T})\\
&+(\frac{3}{2}C_{7}-\frac{1}{3}C_{3}+\frac{5}{3}C_{9}-C_{4}+\frac{1}{2}C_{8}+C_{10})(A^{LL}_{S}+A^{LL}_{T})\\
&+(\frac{1}{2}C_{9}-C_{3}-\frac{1}{2}C_{10}-2C_{4})M^{LL}_{T}-(\frac{1}{2}C_{8}+2C_{6})M^{SP}_{T}\\
&+(\frac{1}{6}C_{7}-\frac{1}{3}C_{5}+\frac{1}{2}C_{8}-C_{6})(A^{SP}_{S}+A^{SP}_{T}+F^{SP}_{S})\\
&+(\frac{1}{2}C_{7}-C_{5})(W^{LR}_{T}+W^{LR}_{S}+M^{LR}_{T}+M^{LR}_{S})]\},
\end{split}
\end{equation}

\begin{equation}
\begin{split}
{\cal A}(B^{0} \rightarrow f^{s}_{2}a^{0}_{0}[ \rightarrow \pi^{0}\eta, K^{+}K^{-}])=
&-\frac{G_{F}}{2}V^{*}_{tb}V_{td}[(-C_{4}+\frac{1}{2}C_{10})M^{LL}_{T}+(-C_{6}+\frac{1}{2}C_{8})M^{SP}_{T}],\\
\end{split}
\end{equation}
where superscripts $LL$, $LR$, $SP$ refer to the the contributions from $(V-A)\otimes(V-A)$, $(V-A)\otimes(V+A)$ and $(S-P)\otimes(S+P)$ operators, respectively. The explicit formulas for the factorizable emission (annihilation) contributions $F_{S}(A_{S})$ shown in Fig.~\ref{fig:figure1}$a1,b1$,$(e1,f1)$, and the nonfactorizable emission (annihilation) contributions $M_{S}(W_{S})$ from Fig.~\ref{fig:figure1}$c1,d1$,$(g1,h1)$ can be obtained in Appendix, as $A_{T}, M_{T}, W_{T}$.

As discussed the quark flavor basis of the mixing, the decay amplitudes for the physical states are then

\begin{equation}
\begin{split}
{\cal A}(B^{0} \rightarrow K^{*}_{2} f_{0}[ \rightarrow \pi^{+}\pi^{-}, K^{+}K^{-}])={\cal A}(B^{0} \rightarrow K^{*}_{2} f_{n}[ \rightarrow \pi^{+}\pi^{-}, K^{+}K^{-}])\sin\theta+{\cal A}(B^{0} \rightarrow K^{*}_{2} f_{s}[ \rightarrow \pi^{+}\pi^{-}, K^{+}K^{-}])\cos\theta
\end{split}
\end{equation}

\begin{equation}
\begin{split}
{\cal A}(B^{0} \rightarrow K^{*}_{2} \sigma[ \rightarrow \pi^{+}\pi^{-}])={\cal A}(B^{0} \rightarrow K^{*}_{2} f_{n}[ \rightarrow \pi^{+}\pi^{-}])\cos\theta-{\cal A}(B^{0} \rightarrow K^{*}_{2} f_{s}[ \rightarrow \pi^{+}\pi^{-}])\sin\theta
\end{split}
\end{equation}

\begin{equation}
\begin{split}
{\cal A}(B^{0} \rightarrow f^{'}_{2}a^{0}_{0}[ \rightarrow \pi^{0}\eta, K^{+}K^{-}])={\cal A}(B^{0} \rightarrow f^{n}_{2}a^{0}_{0}[ \rightarrow \pi^{0}\eta, K^{+}K^{-}])\sin\phi-{\cal A}(B^{0} \rightarrow f^{s}_{2}a^{0}_{0}[ \rightarrow \pi^{0}\eta, K^{+}K^{-}])\cos\phi
\end{split}
\end{equation}

\begin{equation}
\begin{split}
{\cal A}(B^{0} \rightarrow f_{2}a^{0}_{0}[ \rightarrow \pi^{0}\eta, K^{+}K^{-}])={\cal A}(B^{0} \rightarrow f^{n}_{2}a^{0}_{0}[ \rightarrow \pi^{0}\eta, K^{+}K^{-}])\cos\phi+{\cal A}(B^{0} \rightarrow f^{s}_{2}a^{0}_{0}[ \rightarrow \pi^{0}\eta, K^{+}K^{-}])\sin\phi
\end{split}
\end{equation}

\begin{equation}
\begin{split}
{\cal A}(B^{0} \rightarrow f^{'}_{2} f_{0}[ \rightarrow \pi^{+}\pi^{-}, K^{+}K^{-}])=&{\cal A}(B^{0} \rightarrow f^{n}_{2} f_{n}[ \rightarrow \pi^{+}\pi^{-}, K^{+}K^{-}])\sin\phi\sin\theta\\
+&{\cal A}(B^{0} \rightarrow f^{n}_{2} f_{s}[ \rightarrow \pi^{+}\pi^{-}, K^{+}K^{-}])\sin\phi\cos\theta\\
-&{\cal A}(B^{0} \rightarrow f^{s}_{2} f_{n}[ \rightarrow \pi^{+}\pi^{-}, K^{+}K^{-}])\cos\phi\sin\theta\\
-&{\cal A}(B^{0} \rightarrow f^{s}_{2} f_{s}[ \rightarrow \pi^{+}\pi^{-}, K^{+}K^{-}])\cos\phi\cos\theta
\end{split}
\end{equation}

\begin{equation}
\begin{split}
{\cal A}(B^{0} \rightarrow f_{2} f_{0}[ \rightarrow \pi^{+}\pi^{-}, K^{+}K^{-}])=&{\cal A}(B^{0} \rightarrow f^{n}_{2} f_{n}[ \rightarrow \pi^{+}\pi^{-}, K^{+}K^{-}])\cos\phi\sin\theta\\
+&{\cal A}(B^{0} \rightarrow f^{n}_{2} f_{s}[ \rightarrow \pi^{+}\pi^{-}, K^{+}K^{-}])\cos\phi\cos\theta\\
+&{\cal A}(B^{0} \rightarrow f^{s}_{2} f_{n}[ \rightarrow \pi^{+}\pi^{-}, K^{+}K^{-}])\sin\phi\sin\theta\\
+&{\cal A}(B^{0} \rightarrow f^{s}_{2} f_{s}[ \rightarrow \pi^{+}\pi^{-}, K^{+}K^{-}])\sin\phi\cos\theta
\end{split}
\end{equation}

\begin{equation}
\begin{split}
{\cal A}(B^{0} \rightarrow f^{'}_{2} \sigma[ \rightarrow \pi^{+}\pi^{-}])=&{\cal A}(B^{0} \rightarrow f^{n}_{2} f_{n}[ \rightarrow \pi^{+}\pi^{-}])\sin\phi\cos\theta-{\cal A}(B^{0} \rightarrow f^{n}_{2} f_{s}[ \rightarrow \pi^{+}\pi^{-}])\sin\phi\sin\theta\\
-&{\cal A}(B^{0} \rightarrow f^{s}_{2} f_{n}[ \rightarrow \pi^{+}\pi^{-}])\cos\phi\cos\theta+{\cal A}(B^{0} \rightarrow f^{s}_{2} f_{s}[ \rightarrow \pi^{+}\pi^{-}])\cos\phi\sin\theta
\end{split}
\end{equation}

\begin{equation}
\begin{split}
{\cal A}(B^{0} \rightarrow f_{2} \sigma[ \rightarrow \pi^{+}\pi^{-}])=&{\cal A}(B^{0} \rightarrow f^{n}_{2} f_{n}[ \rightarrow \pi^{+}\pi^{-}])\cos\phi\cos\theta-{\cal A}(B^{0} \rightarrow f^{n}_{2} f_{s}[ \rightarrow \pi^{+}\pi^{-}])\cos\phi\sin\theta\\
+&{\cal A}(B^{0} \rightarrow f^{s}_{2} f_{n}[ \rightarrow \pi^{+}\pi^{-}])\sin\phi\cos\theta-{\cal A}(B^{0} \rightarrow f^{s}_{2} f_{s}[ \rightarrow \pi^{+}\pi^{-}])\sin\phi\sin\theta
\end{split}
\end{equation}


\section{Numerical Results And Discussions}\label{sec:numer}

With the decay amplitudes $\cal A$, the differential branching ratio for the $B^{0}\rightarrow TS[ \rightarrow P_{1}P_{2}]$ decays can be taken as

\begin{equation}
\frac{d\cal B}{d\omega}=\frac{\tau\omega\mid\vec{p}_{1}\mid\mid\vec{p}_{3}\mid}{32\pi^{3}M_{B}^{3}}\mid \overline{\cal A}\mid^{2},
\end{equation}
where $\tau$ is the $B$ meson lifetime, $\mid\vec{p}_{1}\mid$ and $\mid\vec{p}_{3}\mid$ respectively denote the magnitudes of momentum for one of the $P_{1}P_{2}$ meson pairs and the tensor meson $T$

\begin{equation}
\begin{split}
\mid\vec{p}_{1}\mid=\frac{\lambda^{1/2}(\omega^{2}, m_{P_{1}}^{2},m_{P_{2}}^{2})}{2\omega},\\
\mid\vec{p}_{3}\mid=\frac{\lambda^{1/2}(M_{B}^{2}, m_{T}^{2},\omega^{2})}{2\omega}
\end{split}
\end{equation}
with the K\"{a}ll\'{e}n function $\lambda(a,b,c)=a^{2}+b^{2}+c^{2}-2(ab+ac+bc)$.

Meanwhile, the direct $CP$ asymmetries of these decays can be defined as
\begin{equation}
\begin{split}
\mathcal{A}_{C P}=\frac{\mathcal{B}(\bar{B} \rightarrow \bar{f})-\mathcal{B}(B \rightarrow f)}{\mathcal{B}(\bar{B} \rightarrow \bar{f})+\mathcal{B}(B \rightarrow f)}\\
\end{split}
\end{equation}
where $ \overline f $ is the $CP$ conjugate state of $ f $. Obviously, both the branching ratios and the direct $CP$ asymmetries are related to the mixing angle.

In Table \ref{tab}, we present the input parameters used in the calculations, including the masses and decay constants of the mesons, the lifetime of the $B^{0}$ meson, and the Wolfenstein parameters of the CKM matrix elements~\cite{Yan:2023yvx,J:2022wa,ParticleDataGroup:2022pth,Dai:2023knn}.

\begin{table}[htbp]
\centering
\caption{Input parameters of the $B^{0} \rightarrow TS[ \rightarrow P_{1}P_{2}]$ decays}
\label{tab}
\begin{tabular*}{\columnwidth}{@{\extracolsep{\fill}}llllll@{}}
\hline
\hline
\\
&$M_{B}=5.28$ {\rm GeV}       &$m_{b}=4.2$ {\rm GeV}     &$f_{B}=0.19\pm0.02$ {\rm GeV}    &$\tau=1.519$ {\rm ps}\\
                                  \\
&$m_{a_{0}}=0.98\pm0.02$ {\rm GeV}&$m_{f_{0}}=0.99\pm0.02$ {\rm GeV} &$m_{\sigma}=0.5$ {\rm GeV}&$\overline{f}_{s}=0.37$ {\rm GeV}\\
                                  \\
&$m_{K^{*0}_{2}}=1.432$ {\rm GeV}&$M_{K^{*\pm}_{2}}=1.427$ {\rm GeV} &$M_{a_{2}}=1.317$ {\rm GeV} &$M_{f^{'}_{2}}=1.517$ {\rm GeV}\\
                                  \\
&$M_{f_{2}}=1.275$ {\rm GeV}  &$f_{a_{2}}=107\pm 6$ {\rm MeV}&$ f_{K^{*}_{2}}=118\pm 5$ {\rm MeV}  	&$f_{f_{2}}=102\pm 6$ {\rm MeV} \\
\\
&$ f_{f^{'}_{2}}=126\pm 4$ {\rm MeV} &$f^{T}_{a_{2}}=105\pm 21$ {\rm MeV}&$ f^{T}_{K^{*}_{2}}=77\pm 14$ {\rm MeV}  &$f^{T}_{f_{2}}=117\pm 25$ {\rm MeV}\\
\\
&$ f^{T}_{f^{'}_{2}}=65\pm 12$ {\rm MeV}
&$m_{\pi^{\pm}}=0.14$ {\rm GeV}   &$m_{\pi^{0}}=0.135$ {\rm GeV}  &$m_{K^{\pm}}=0.494$ {\rm GeV}\\
                                  \\
&$m_{K^{0}}=0.498$ {\rm GeV} &$m_{\eta}=0.548$ {\rm GeV} \\
                                  \\
&$\lambda=0.22650$  &$\emph{A}=0.790$   &$\bar{\rho}=0.141$   &$\bar{\eta}=0.357$\\
                                  \\
\hline
\hline
\end{tabular*}
\end{table}

By using the helicity amplitudes and the input parameters, we predict the $CP$ averaged branching fractions of the $B^{0} \rightarrow TS[ \rightarrow P_{1}P_{2}]$ decays in the pQCD approach, and make some comments on the results. There are still many uncertainties in our calculation results. We primarily consider three types of errors. The first error is the decay constants of tensor mesons, the second is the shape parameter of $B$ meson $\omega_{b}=0.40\pm0.04$ GeV, and the third comes from the Gegenbauer coefficient $a_{2}=0.3\pm0.1$ for $a_{0}$ and $a_{2}=0.3\pm0.2$ for $f_{0}$. We have neglected the uncertainties caused by the Wolfenstein parameters $\lambda, A, \rho, \eta$ because they are typically small. In considered branching ratios, the major uncertainties stem from the shape parameter $\omega_{b}$ of the $B$ meson's wave function and the Gegenbauer coefficient $a_{2}$. It is well known that the S-wave two-meson distribution amplitudes are not well determined, this leads to significant errors in the Gegenbauer coefficient $a_{2}$. The rigorous theoretical calculation for nonresonance contribution in the context of the PQCD framework is still absent, and the comparison between experiment measurements and theoretical predictions is still challenging. We hope that future research can further explore more precise description methods. It is worth mentioning that the $\omega_{b}$ plays vastly different roles in various decay modes. In Table \ref{a}, the branching ratios of $B^{0} \rightarrow K^{*}_{2}a^{0}_{0}[ \rightarrow \pi^{0}\eta,K^{+}K^{-}]$ exhibit significantly higher sensitivity to the shape parameter $\omega_{b}$ than those of $B^{0} \rightarrow K^{*+}_{2}a^{-}_{0}[ \rightarrow \pi^{-}\eta,K^{-}K^{0}]$. Firstly, the contribution of the factorizable annihilation diagrams is unrelated to $\omega_{b}$. Meanwhile, the $B^{0} \rightarrow K^{*}_{2}a^{0}_{0}[ \rightarrow \pi^{0}\eta,K^{+}K^{-}]$ and $B^{0} \rightarrow K^{*+}_{2}a^{-}_{0}[ \rightarrow \pi^{-}\eta,K^{-}K^{0}]$ are primarily dominated by tree operator contributions, and the nonfactorized emission diagrams will generate significant uncertainty from $\omega_{b}$. However, the latter is suppressed by the coefficient $C1$, resulting in a relatively smaller impact from $\omega_{b}$.

\begin{figure}[htbp]
	\centering
	\begin{tabular}{l}
		\includegraphics[width=0.5\textwidth]{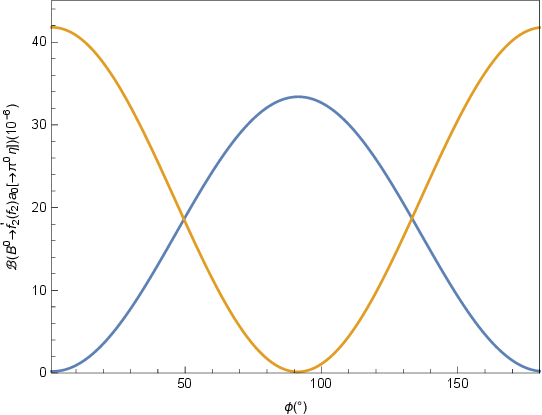}
	\end{tabular}
	\caption {The branching ratio of $B^{0} \rightarrow f^{'}_{2}(f_{2})a^{0}_{0}[ \rightarrow \pi^{0}\eta]$ decay with a variant of the mixing angle $\phi$.}
	\label{fig2}
\end{figure}

\begin{figure}[htbp]
	\centering
	\begin{tabular}{l}
		\includegraphics[width=0.5\textwidth]{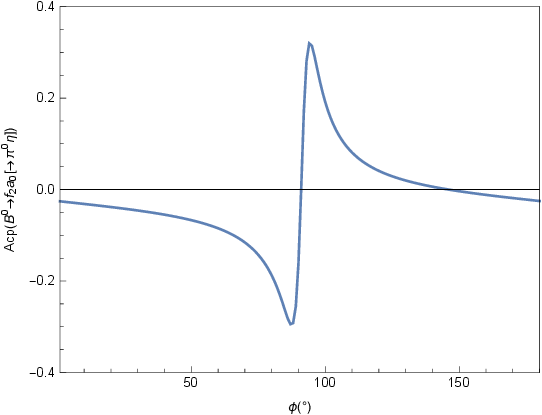}
		\includegraphics[width=0.5\textwidth]{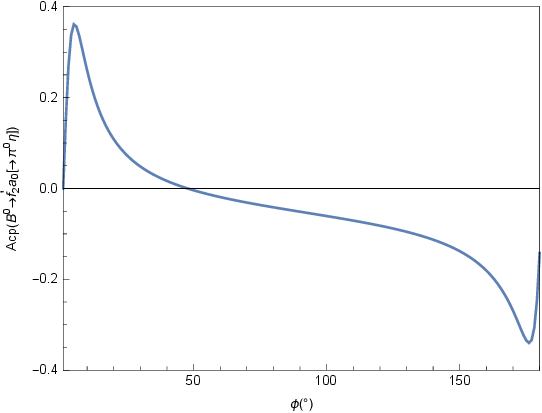}
	\end{tabular}
	\caption {The direct $CP$ asymmetries of $B^{0} \rightarrow f^{'}_{2}(f_{2})a^{0}_{0}[ \rightarrow \pi^{0}\eta]$ with a variant of the mixing angle $\phi$.}
	\label{fig3}
\end{figure}

In Fig.\ref{fig2} and Fig.\ref{fig3}, by setting the mixing angle $\phi$ to be a free parameter, we plot the variation of the branching fractions and direct $CP$ asymmetries of decays $B^{0} \rightarrow f^{'}_{2}a^{0}_{0}[ \rightarrow \pi^{0}\eta]$ and $B^{0} \rightarrow f_{2}a^{0}_{0}[ \rightarrow \pi^{0}\eta]$ with the angle $\phi$, respectively. In Fig.\ref{fig2}, the pattern of change in the decay branching ratio of $B^{0} \rightarrow f^{'}_{2}a^{0}_{0}[ \rightarrow \pi^{0}\eta]$, indicated by the blue line, agrees with the image of the sine function, while the change in the decay branching ratio of $B^{0} \rightarrow f_{2}a^{0}_{0}[ \rightarrow \pi^{0}\eta]$ with the mixing angle, indicated by the yellow line, satisfies the law of the cosine function. The detailed discussions about the mixing angle $\phi$ could be found in Refs\cite{Zou:2012td,ParticleDataGroup:2022pth,Cheng:2011fk,Li:2000zb,Li:2018lbd}. In Table \ref{a}, we employ the most recent updated value $9 ^{\circ}$. It is known to us that the direct $CP$ asymmetry is proportional to the interference between contributions from the tree and penguin operators. In Fig.\ref{fig2} and Fig.\ref{fig3}, when the mixing angle $\phi=0^{\circ}$, $f^{'}_{2}$ is regarded as the pure $s\overline{s}$, the branching ratio of $B^{0} \rightarrow f^{'}_{2}a^{0}_{0}[ \rightarrow \pi^{0}\eta]$ is about $2.11\times10^{-7}$, the direct $CP$ asymmetries are zero since it is pure penguin processes, and the both of value will increase a lot after considering the mixing of $\frac{1}{\sqrt{2}}(u\overline{u}+d\overline{d})$. Whereas for $B^{0} \rightarrow f_{2}a^{0}_{0}[ \rightarrow \pi^{0}\eta]$ decay, the branching ratio is $4.18\times10^{-5}$ with $\phi=0^{\circ}$, the direct $CP$ asymmetries are $-2.57\%$ and the values will have little change after considering the mixing of $s\overline{s}$. Moreover, in Table \ref{a}, we find that the branching fractions of $B^{0} \rightarrow f_{2}a^{0}_{0}[ \rightarrow \pi^{0}\eta]$ are 2 orders of magnitude larger than that of $B^{0} \rightarrow f^{'}_{2}a^{0}_{0}[ \rightarrow \pi^{0}\eta]$. Therefore, $f^{n}_{2}=\frac{1}{\sqrt{2}}(u\overline{u}+d\overline{d})$ makes the dominant contribution in the decays, and $B^{0} \rightarrow f^{'}_{2}a^{0}_{0}[ \rightarrow \pi^{0}\eta]$ are sensitive to the the interference between $f^{n}_{2}$ and $f^{s}_{2}$.

In Table \ref{a}, we present the branching fractions and direct $CP$ asymmetries of the $B^{0} \rightarrow Ta_{0}[ \rightarrow \pi\eta, K\overline{K}]$ decays. Meanwhile, we can find the branching ratios of the ${B}^{0} \rightarrow Ta_{0}[ \rightarrow \pi\eta]$ decays are much larger than that of the ${B}^{0} \rightarrow Ta_{0}[ \rightarrow K\overline{K}]a_{0}$ decays, which can be explained by the fact that the phase space for $K\overline{K}$ is suppressed. At the same time, the authors also concluded that the branching fractions of the $\pi\eta$ channel were five times larger than that of the $K\overline{K}$ channel with the resonance $a_{0}(980)$ in Ref.~\cite{J:2022wa}. Next, we will use our calculations to investigate the value of $\Gamma(a_{0} \rightarrow K^{+}K^{-})/\Gamma(a_{0} \rightarrow \pi^{0}\eta)$ with the narrow-width approximation.

\begin{table}[htbp]
	\centering
	\caption{$CP$-averaged branching fractions and the direct $CP$ asymmetries for the $B^{0} \rightarrow Ta_{0}[ \rightarrow \pi\eta, K\overline{K}]$ decays in the pQCD approach.}
	\label{a}
	\begin{tabular*}{\columnwidth}{@{\extracolsep{\fill}}llllll@{}}
		\hline
		\hline\\
		&Decay modes         & ${\cal B}$     & ${\cal A}_{C P}$ \\
		\hline
		\\
		&$B^{0} \rightarrow K^{*}_{2}(1430)a^{0}_{0}[ \rightarrow \pi^{0}\eta]$  & $ 1.40^{+0.10+1.32+1.23}_{-0.07-0.64-0.81}\times10^{-6}$   & $-5.17^{+4.10+10.76+29.98}_{-4.61-6.49-13.33}\%$   \\
		\\
		&$B^{0} \rightarrow K^{*}_{2}(1430)a^{0}_{0}[ \rightarrow K^{+}K^{-}]$  & $ 1.64^{+0.10+1.51+1.40}_{-0.07-0.73-0.93}\times10^{-7}$   & $-3.73^{+4.34+9.31+29.27}_{-4.70-6.61-13.44}\%$   \\
		\\
		&$B^{0} \rightarrow K^{*+}_{2}(1430)a^{-}_{0}[ \rightarrow \pi^{-}\eta]$  & $ 2.78^{+0.34+0.50+1.03}_{-0.32-0.29-0.76}\times10^{-7}$   & $-35.80^{+1.66+0.93+2.18}_{-1.77-0.96-2.75}\%$   \\
		\\
		&$B^{0} \rightarrow K^{*+}_{2}(1430)a^{-}_{0}[ \rightarrow K^{-}K^{0}]$  & $ 5.25^{+0.70+1.02+2.12}_{-0.66-0.63-1.55}\times10^{-8}$   & $-41.82^{+1.71+2.36+2.22}_{-1.96-0.96-0.78}\%$   \\
		\\
		&$B^{0} \rightarrow a^{0}_{2}(1320)a^{0}_{0}[ \rightarrow \pi^{0}\eta]$  & $ 2.37^{+1.33+0.68+3.35}_{-1.01-0.35-1.88}\times10^{-6}$   & $-2.78^{+0.87+0.71+3.69}_{-1.44-0.11-12.63}\%$   \\
		\\
		&$B^{0} \rightarrow a^{0}_{2}(1320)a^{0}_{0}[ \rightarrow K^{+}K^{-}]$  & $ 3.33^{+1.48+1.45+4.54}_{-1.15-0.71-2.59}\times10^{-7}$   & $-3.90^{+0.97+0.87+3.75}_{-1.41-0.22-12.08}\%$   \\
		\\
		&$B^{0} \rightarrow a^{+}_{2}(1320)a^{-}_{0}[ \rightarrow \pi^{-}\eta]$  & $ 5.43^{+0.94+2.17+4.02}_{-0.86-1.41-2.88}\times10^{-6}$   & $11.57^{+0.52+1.44+4.07}_{-0.60-1.37-1.99}\%$   \\
		\\
		&$B^{0} \rightarrow a^{+}_{2}(1320)a^{-}_{0}[ \rightarrow K^{-}K^{0}]$  & $ 1.12^{+0.21+0.44+0.82}_{-0.18-0.28-0.58}\times10^{-6}$   & $11.82^{+0.62+1.34+3.93}_{-0.72-1.32-2.00}\%$   \\
		\\
		&$B^{0} \rightarrow a^{-}_{2}(1320)a^{+}_{0}[ \rightarrow \pi^{+}\eta]$  & $ 5.59^{+1.52+4.85+5.61}_{-0.21-3.60-3.18}\times10^{-7}$   & $13.84^{+12.93+0.29+17.48}_{-17.41-16.42-3.69}\%$   \\
		\\
		&$B^{0} \rightarrow a^{-}_{2}(1320)a^{+}_{0}[ \rightarrow K^{+}K^{0}]$  & $ 1.71^{+0.06+1.08+1.69}_{-0.01-0.87-1.03}\times10^{-7}$   & $-9.61^{+12.93+8.47+4.90}_{-8.99-18.54-2.83}\%$   \\
		\\
		&$B^{0} \rightarrow f^{'}_{2}(1525)a^{0}_{0}[ \rightarrow \pi^{0}\eta]$  & $ 7.76^{+0.75+6.34+7.14}_{-0.67-3.43-4.76}\times10^{-7}$   & $28.53^{+3.55+2.83+1.51}_{-3.27-2.75-5.85}\%$   \\
		\\
		&$B^{0} \rightarrow f^{'}_{2}(1525)a^{0}_{0}[ \rightarrow K^{+}K^{-}]$  & $ 8.21^{+0.78+6.79+7.59}_{-0.71-3.61-5.02}\times10^{-8}$   & $28.06^{+3.43+2.07+2.70}_{-3.22-2.22-9.46}\%$   \\
		\\
		&$B^{0} \rightarrow f_{2}(1270)a^{0}_{0}[ \rightarrow \pi^{0}\eta]$  & $ 4.10^{+0.83+3.18+3.27}_{-0.72-1.76-2.30}\times10^{-5}$   & $-3.08^{+1.60+0.53+1.56}_{-1.97-0.67-0.81}\%$   \\
		\\
		&$B^{0} \rightarrow f_{2}(1270)a^{0}_{0}[ \rightarrow K^{+}K^{-}]$  & $ 4.55^{+0.93+3.50+3.58}_{-0.80-1.96-2.53}\times10^{-6}$   & $-2.76^{+1.59+0.33+1.94}_{-2.00-0.44-1.00}\%$   \\
		\hline
		\hline
	\end{tabular*}
\end{table}

When the narrow-width approximation is considered, the branching ratio of the quasi-two-body decay can be written as
\begin{equation}
{\cal B}(B \rightarrow M_{1}(R \rightarrow )M_{2}M_{3})\simeq{\cal B}(B \rightarrow M_{1}R)\times{\cal B}(R \rightarrow M_{2}M_{3}),
\end{equation}
with the resonance $R$. We can define a ratio ${\cal R}_{1}$ as

\begin{equation}
\begin{split}
{\cal R}_{1}=\frac{{\Gamma}(a_{0} \rightarrow K^{+}K^{-})}{{\Gamma}(a_{0} \rightarrow \pi^{0}\eta)}=\frac{{\cal B}({B}^{0} \rightarrow Ta^{0}_{0})\times{\cal B}(a^{0}_{0} \rightarrow K^{+}K^{-})}{{\cal B}({B}^{0} \rightarrow Ta^{0}_{0})\times{\cal B}(a^{0}_{0} \rightarrow \pi^{0}\eta)}\simeq\frac{{\cal B}({B}^{0} \rightarrow Ta^{0}_{0}[ \rightarrow K^{+}K^{-}])}{{\cal B}({B}^{0} \rightarrow Ta^{0}_{0}[ \rightarrow \pi^{0}\eta])}.
\end{split}
\end{equation}

After considering the isospin relation $\Gamma(a_{0} \rightarrow K^{+}K^{-})=\Gamma(a_{0} \rightarrow K\overline{K})/2$, we obtain the relative partial decay width

\begin{equation}
\frac{\Gamma(a_0 \rightarrow K \bar{K})}{\Gamma(a_0 \rightarrow \pi^{0} \eta)} \approx \begin{cases}0.23  & \text { $K^{*}_{2}(1430)$ } \\ 0.28  & \text { $a^{0}_{2}(1320)$ }\\ 0.21  & \text { $f^{'}_{2}(1525)$ }\\ 0.22  & \text { $f_{2}(1270)$ }\end{cases}
\end{equation}
respectively. The OBELIX Collaboration acquired the ratio $\Gamma(a_{0} \rightarrow K\overline{K}) / \Gamma(a_{0} \rightarrow \pi^{0}\eta) = 0.57 \pm 0.16$ using data from $\pi^{+}\pi^{-}\pi^{0}$, $K^{+}K^{-}\pi^{0}$, and $K^{\pm}K^{0}{S}\pi^{\mp}$\cite{M:2003bgg}. In Ref.\cite{Abele:1998hls}, the authors found the branching ratio for ${\cal B}(p\overline{p} \rightarrow a{0}(980)\pi \rightarrow K\overline{K}\pi)=5.92^{+0.46}_{-1.01} \times 10^{-4}$, and combined this with ${\cal B}(p\overline{p} \rightarrow a_{0}(980)\pi\rightarrow \pi^{0}\pi^{0}\eta) = (2.61 \pm 0.48) \times 10^{-4}$\cite{C:1994yn}, yielding $\Gamma(a_{0} \rightarrow K\overline{K}) / \Gamma(a_{0} \rightarrow \pi\eta) = 0.23 \pm 0.05$. The WA102 Collaboration reported $\Gamma(a_{0} \rightarrow K\overline{K}) / \Gamma(a_{0} \rightarrow \pi\eta) = 0.166 \pm 0.01 \pm 0.02$ from $f_{1}(1285)$ decays\cite{Barberis:1998zzb}. From this, it can be seen that for different decay processes, the analysis of the partial decay width ratio $\Gamma(a_{0} \rightarrow K\overline{K}) / \Gamma(a_{0} \rightarrow \pi\eta)$ may yield some differences\cite{R:2019ngy,J:2022wa,Wang:2024enc}. The Particle Data Group provides an average ratio of $\Gamma(a_{0} \rightarrow K\overline{K}) / \Gamma(a_{0} \rightarrow \pi\eta) = 0.183 \pm 0.024$. Our results are slightly lager than this average but are consistent with the data from Ref.\cite{Abele:1998hls} within the errors.

\begin{figure}[htbp]
	\centering
	\begin{tabular}{l}
		\includegraphics[width=0.5\textwidth]{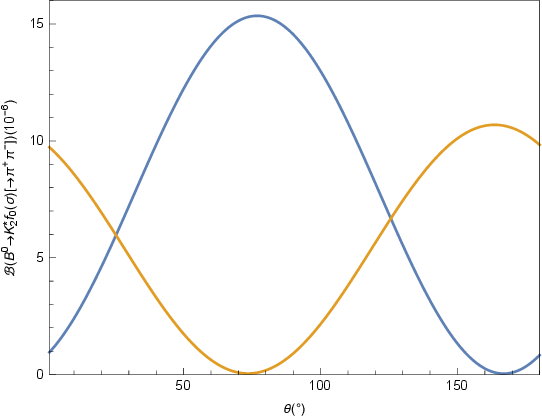}
		\includegraphics[width=0.5\textwidth]{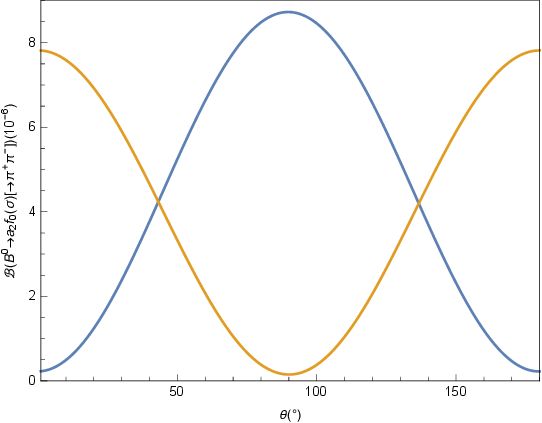}\\
	\end{tabular}
	\caption {The branching ratio of $B^{0} \rightarrow K^{*}_{2}f_{0}(\sigma)[ \rightarrow \pi^{+}\pi^{-}]$ and $B^{0} \rightarrow a^{0}_{2}f_{0}(\sigma)[ \rightarrow \pi^{+}\pi^{-}]$ decay with variant of the mixing angle $\theta$.}
	\label{fig4}
\end{figure}

\begin{figure}[htbp]
	\centering
	\begin{tabular}{l}
		\includegraphics[width=0.5\textwidth]{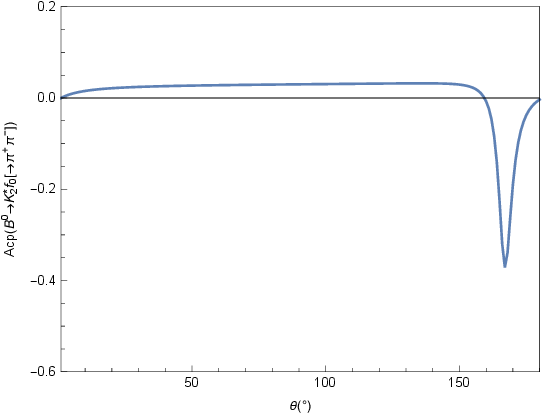}
		\includegraphics[width=0.5\textwidth]{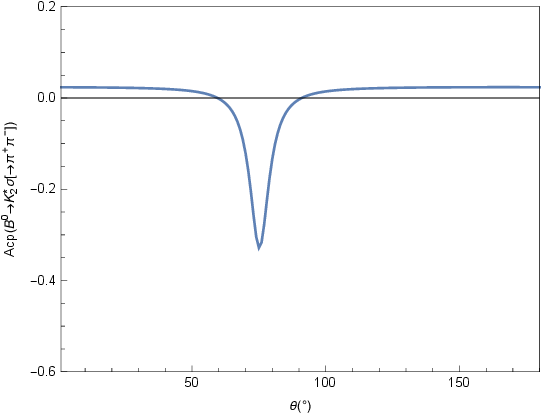}\\
		\includegraphics[width=0.5\textwidth]{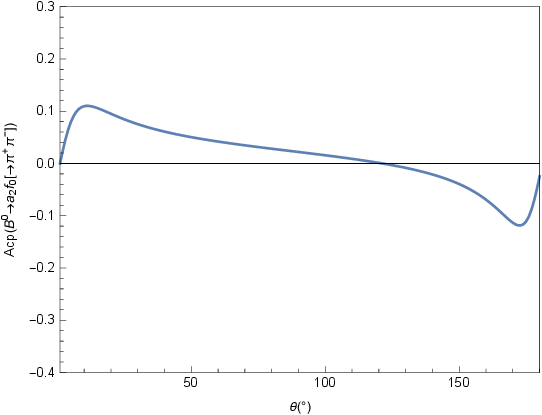}
		\includegraphics[width=0.5\textwidth]{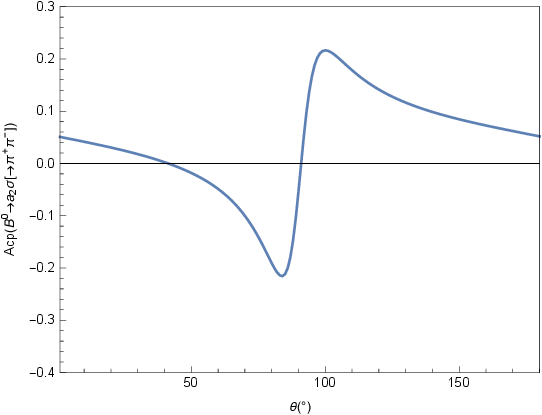}\\
	\end{tabular}
	\caption {The direct $CP$ asymmetries of $B^{0} \rightarrow K^{*}_{2}f_{0}(\sigma)[ \rightarrow \pi^{+}\pi^{-}]$ and $B^{0} \rightarrow a^{0}_{2}f_{0}(\sigma)[ \rightarrow \pi^{+}\pi^{-}]$ with variant of the mixing angle $\theta$.}
	\label{fig5}
\end{figure}

The mixing angle $\theta$ is introduced into the $f_{0}-\sigma$ mixing mechanism, which has not been determined precisely by current experimental measurements, and is suggested to be in the wide ranges of $25^{\circ}<\theta<40^{\circ}$\cite{Niu:2021pcx,Liu:2019ymi,Cheng:2023knr,Yang:2022jog} and $135^{\circ}<\theta<165^{\circ}$\cite{Yan:2023yvx,Cheng:2002ai,Gokalp:2004ny,L:2021vb,Li:2019jlp}. In Figure.\ref{fig4} and Figure.\ref{fig5}, by setting the mixing angle $\theta$ to be a free parameter, we plot the variation of the branching fractions and direct $CP$ asymmetries of decays $B^{0} \rightarrow K^{*}_{2}f_{0}(\sigma)[ \rightarrow \pi^{+}\pi^{-}]$ and $B^{0} \rightarrow a^{0}_{2}f_{0}(\sigma)[ \rightarrow \pi^{+}\pi^{-}]$ with the angle $\theta$, respectively. In Figure.\ref{fig4}, the pattern of change in the decay branching ratio of $B^{0} \rightarrow K^{*}_{2}f_{0}[ \rightarrow \pi^{+}\pi^{-}]$ and $B^{0} \rightarrow a^{0}_{2}f_{0}[ \rightarrow \pi^{+}\pi^{-}]$ , indicated by the blue line, the change in the decay branching ratio of $B^{0} \rightarrow K^{*}_{2}\sigma[ \rightarrow \pi^{+}\pi^{-}]$ and $B^{0} \rightarrow a^{0}_{2}\sigma[ \rightarrow \pi^{+}\pi^{-}]$ with the mixing angle, indicated by the yellow line. We find graphically that the contribution from the $f_{n}=\frac{1}{\sqrt{2}}(u\overline{u}+d\overline{d})$ component is dominant. For these modes, both $f_{n}$ and $f_{s}$ will contribute, but the relative sign of the $f_{s}$ state with respect to the $f_{n}$ is negative for the $\sigma$ and positive for the $f_{0}$, which leads to a destructive interference between $f_{n}$ and $f_{s}$ for $B^{0} \rightarrow K^{*}_{2}(a^{0}_{2})\sigma[ \rightarrow \pi^{+}\pi^{-}]$, but a constructive interference for $B^{0} \rightarrow K^{*}_{2}(a^{0}_{2})f_{0}[ \rightarrow \pi^{+}\pi^{-}]$.

As we all known, both strong and weak phases are the necessary conditions for direct $CP$ asymmetry. In Figure.\ref{fig5}, we can observe when $\theta=0^{\circ}$, the direct CP asymmetries of decay $B^{0} \rightarrow K^{*}_{2}f_{0}[ \rightarrow \pi^{+}\pi^{-}]$ is zero, since the decay is induced by $b \rightarrow ss\bar{s}$ transition. In the Wolfenstein parametrization of CKM matrix, there is no weak phase in this transition, which is a pure penguin process, the direct $CP$ asymmetries is zero. For $B^{0} \rightarrow K^{*}_{2}\sigma[ \rightarrow \pi^{+}\pi^{-}]$ that is induced by $b \rightarrow sq\bar{q} (q = u, d)$, the direct $CP$ asymmetries decay is less than $5\%$, because $\left|V_{u s} V_{u b}\right| \ll\left|V_{t s} V_{t b}\right|$. Without the influence of mixing angles, the decay process $B^{0} \rightarrow a^{0}_{2}f_{0}[ \rightarrow \pi^{+}\pi^{-}]$ solely receives penguin contributions from the $c1$ and $d1$ nonfactorizable emission diagrams, resulting in a direct $CP$ asymmetry of 0. Similarly, in the decay $B^{0} \rightarrow a^{0}_{2}\sigma[ \rightarrow \pi^{+}\pi^{-}]$, the interference between the tree and penguin contributions is minimal, yielding small direct $CP$ asymmetries with $5.08\%$ . However, when we incorporate mixing effects, the $f_{n}$ term introduces tree contributions, thereby rendering the direct $CP$ asymmetries of both $B^{0} \rightarrow K^{*}_{2}f_{0}[ \rightarrow \pi^{+}\pi^{-}]$ and $B^{0} \rightarrow a^{0}_{2}f_{0}[ \rightarrow \pi^{+}\pi^{-}]$ are no longer zero in this work. Therefore, if the two-quark structure will be confirmed, the branching fractions and $CP$ asymmetries can be used to determine the mixing angle $\theta$ and study the nature of scalar particles.

\begin{table}[htbp]
	\centering
	\caption{Branching ratios and the direct $CP$ asymmetries for the ${B}^{0} \rightarrow Tf_{0}[ \rightarrow \pi^{+}\pi^{-},K^{+}K^{-}]$ and ${B}^{0} \rightarrow T\sigma[ \rightarrow \pi^{+}\pi^{-}]\sigma$ decays in the pQCD approach with the $f_{0}-\sigma$ mixing angle $\theta=135^{\circ}$.}
	\label{f}
	\begin{tabular*}{\columnwidth}{@{\extracolsep{\fill}}lllll@{}}
		\hline
		\hline
        &Decay modes         & ${\cal B}$     & ${\cal A}_{C P}$ \\
		\hline
		\\
		&$B^{0} \rightarrow K^{*}_{2}(1430)f_{0}[ \rightarrow \pi^ {+}\pi^{-}]$  & $ 4.30^{+0.40+2.71+5.52}_{-0.34-1.57-3.21}\times10^{-6}$   & $3.21^{+2.95+2.30+3.02}_{-2.59-1.41-5.99}\%$   \\
		\\
		&$B^{0} \rightarrow K^{*}_{2}(1430)f_{0}[ \rightarrow K^ {+}K^{-}]$      & $ 1.18^{+0.11+0.75+1.46}_{-0.09-0.43-0.86}\times10^{-6}$    & $6.13^{+2.83+1.89+2.66}_{-2.57-0.95-5.52}\%$   \\
		\\
		&$B^{0} \rightarrow K^{*}_{2}(1430)\sigma[ \rightarrow \pi^ {+}\pi^{-}]$  & $ 8.26^{+0.82+8.94+8.67}_{-0.51-4.24-3.88}\times10^{-6}$   & $1.63^{+0.40+0.36+0.20}_{-0.51-1.28-2.32}\%$   \\
		\\
		&$B^{0} \rightarrow a^{0}_{2}(1320)f_{0}[ \rightarrow \pi^ {+}\pi^{-}]$& $ 4.44^{+0.73+5.01+4.88}_{-0.57-2.39-2.09}\times10^{-6}$   & $-1.43^{+1.32+1.31+6.35}_{-1.08-0.38-0.37}\%$   \\
		\\
		&$B^{0} \rightarrow a^{0}_{2}(1320)f_{0}[ \rightarrow K^ {+}K^{-}]$& $ 1.02^{+0.14+1.15+1.14}_{-0.10-0.55-0.47}\times10^{-6}$   & $-1.39^{+1.53+1.62+6.06}_{-1.33-0.53-0.05}\%$   \\
		\\
		&$B^{0} \rightarrow a^{0}_{2}(1320)\sigma[ \rightarrow \pi^ {+}\pi^{-}]$  & $ 3.98^{+0.74+4.31+3.96}_{-0.61-2.09-1.83}\times10^{-6}$   & $10.69^{+2.69+2.20+1.00}_{-2.27-1.82-6.47}\%$   \\
		\\
		&$B^{0} \rightarrow f^{'}_{2}(1525)f_{0}[ \rightarrow \pi^ {+}\pi^{-}]$  & $ 1.10^{+0.04+0.61+1.28}_{-0.04-0.37-0.49}\times10^{-7}$   & $-72.69^{+7.63+2.66+23.46}_{-7.01-2.43-2.47}\%$   \\
		\\
		&$B^{0} \rightarrow f^{'}_{2}(1525)f_{0}[ \rightarrow K^ {+}K^{-}]$  & $ 2.51^{+0.12+1.43+2.96}_{-0.09-0.85-1.08}\times10^{-8}$   & $-72.39^{+8.26+1.43+48.09}_{-7.49-1.08-6.32}\%$   \\
		\\
		&$B^{0} \rightarrow f^{'}_{2}(1525)\sigma[ \rightarrow \pi^ {+}\pi^{-}]$  & $ 3.77^{+0.09+3.78+4.16}_{-0.04-1.86-1.73}\times10^{-8}$   & $-94.94^{+3.21+7.81+50.55}_{-1.49-0.40-0.77}\%$   \\
		\\
    	&$B^{0} \rightarrow f_{2}(1270)f_{0}[ \rightarrow \pi^ {+}\pi^{-}]$  & $ 2.05^{+0.30+1.23+1.47}_{-0.16-0.79-0.77}\times10^{-6}$   & $19.47^{+4.01+0.58+7.18}_{-4.88-0.42-26.68}\%$   \\
    	\\
    	&$B^{0} \rightarrow f_{2}(1270)f_{0}[ \rightarrow K^ {+}K^{-}]$  & $ 5.69^{+1.02+3.62+4.00}_{-0.61-2.28-2.07}\times10^{-7}$   & $14.28^{+4.25+0.28+7.65}_{-4.48-0.15-26.12}\%$   \\
    	\\
	    &$B^{0} \rightarrow f_{2}(1270)\sigma[ \rightarrow \pi^ {+}\pi^{-}]$  & $ 1.34^{+0.13+0.85+1.09}_{-0.02-0.52-0.55}\times10^{-6}$   & $-15.20^{+1.16+4.30+10.99}_{-0.39-3.00-3.91}\%$   \\
		\hline
		\hline
	\end{tabular*}
\end{table}

In the current study, we take $\theta=135^{\circ},\phi=9^{\circ}$ to make a numerical calculation and the Branching ratios and the direct $CP$ asymmetries for the ${B}^{0} \rightarrow Tf_{0}[ \rightarrow \pi^{+}\pi^{-},K^{+}K^{-}]$ and ${B}^{0} \rightarrow T\sigma[ \rightarrow \pi^{+}\pi^{-}]$ decays are presented in Table \ref{f}. We can find that the direct $CP$ asymmetries in the ${B}^{0} \rightarrow Tf_{0}[ \rightarrow \pi^{+}\pi^{-}]$ and ${B}^{0} \rightarrow Tf_{0}[ \rightarrow K^{+}K^{-}]$ decays with the same tensor are comparable. With the reason that the direct $CP$ asymmetry is quantified by the ratio of decay rates, the impact of the scalar timelike form factor parameters and the invariant mass range in the quasi-two-body decay will be basically canceled. We still can find that the local $CP$ asymmetries of the decays $B^{0} \rightarrow K^{*}_{2}f_{0}[ \rightarrow \pi^ {+}\pi^{-},K^ {+}K^{-}]$ are as small as about $5\%$, and the reason is that the tree diagram contributions are both color and CKM elements suppressed. For the decay $B^{0} \rightarrow a^{0}_{2}f_{0}[ \rightarrow \pi^ {+}\pi^{-},K^ {+}K^{-}]$, although the tree operator contributions proportional to the term $(C1 + C2/3)$ are color suppressed, the effects of penguin operators are significantly reduced due to the complete prohibition of factorizable diagrams involving emission tensors by Lorentz invariance, and the heavy suppression of factorizable diagrams for meson pair emission by the vector decay constants of scalars from charge conjugation invariance. Consequently, the $CP$ asymmetries observed in this decay process are indeed very small.

To compare with existing data and further discuss our calculations, we then use the narrow-width approximation to study the $B^{0} \rightarrow K^{*}_{2}f_{0}[ \rightarrow \pi^ {+}\pi^{-},K^{+}K^{-}]$ decays which have the same resonance, we can define a ratio ${\cal R}_{2}$ to describe the relationship between $f_{0} \rightarrow \pi^{+}\pi^{-}$ and $f_{0} \rightarrow K^{+}K^{-}$, which can be given as

\begin{equation}
\begin{split}
{\cal R}_{2}&=\frac{{\cal B}(f_{0} \rightarrow K^{+}K^{-})}{{\cal B}(f_{0} \rightarrow \pi^{+}\pi^{-})}
=\frac{{\cal B}({B}^{0} \rightarrow K^{*}_{2}f_{0})\times{\cal B}(f_{0} \rightarrow K^{+}K^{-})}{{\cal B}({B}^{0} \rightarrow K^{*}_{2}f_{0})\times{\cal B}(f_{0} \rightarrow \pi^{+}\pi^{-})}
\simeq\frac{{\cal B}({B}^{0} \rightarrow K^{*}_{2}f_{0}[ \rightarrow K^{+}K^{-}])}{{\cal B}({B}^{0} \rightarrow K^{*}_{2}f_{0}[ \rightarrow \pi^{+}\pi^{-}])}\approx0.27,
\end{split}
\end{equation}
This ratio can be used to estimate the branching ratios for the $f_{0} \rightarrow \pi^{+}\pi^{-}$ and $f_{0} \rightarrow K^{+}K^{-}$ decays by using the formulas ${\cal B}(f_{0} \rightarrow \pi^{+}\pi^{-})=\frac{2}{4{\cal R}_{2}+3}$ and ${\cal B}(f_{0} \rightarrow K^{+}K^{-})=\frac{2{\cal R}_{2}}{4{\cal R}_{2}+3}$\cite{Fleischer:2011hz}. So we can get

\begin{equation}
\begin{split}
{\cal B}(f_{0} \rightarrow \pi^{+}\pi^{-})\approx0.49,\\
{\cal B}(f_{0} \rightarrow K^{+}K^{-})\approx0.13.
\end{split}
\end{equation}
The BES Collaboration gained the relative branching ratios for the $\psi(2S) \rightarrow \gamma\chi_{c0}$ decays, where $\chi_{c0} \rightarrow f_{0}f_{0} \rightarrow \pi^{+}\pi^{-}\pi^{+}\pi^{-}$ or $\chi_{c0} \rightarrow f_{0}f_{0} \rightarrow \pi^{+}\pi^{-}K^{+}K^{-}$\cite{Ablikim:2004vu,Ablikim:2005rhr}. Meanwhile, the CLEO Collaboration has obtained ${\cal B}(f_{0} \rightarrow K^{+}K^{-})/{\cal B}(f_{0} \rightarrow \pi^{+}\pi^{-})=(25^{+17}_{-11})$\% and extracted ${\cal B}(f_{0} \rightarrow \pi^{+}\pi^{-})=(50^{+7}_{-9})$\% by utilizing BES's findings\cite{Ecklund:2009dvs}. Our calculations are well aligned with CLEO's results and theoretical outcomes\cite{Niu:2021pcx,Liu:2019ymi}. At the same time, based on the narrow width ${\cal B}(f_{0} \rightarrow \pi^{+}\pi^{-})\approx0.49$, we can evaluate ${\cal B}(B^{0} \rightarrow K^{*}_{2}f_{0})=8.78\times10^{-6}$, which agree with the BABAR Collaboration's data\cite{BaBar:2011ryf} and theoretical results performed well in the two-body framework \cite{Li:2019jlp}.

In Table \ref{f}, the considered decays involving $f^{'}_{2}$ are sensitive to the $\phi$, whereas the decays involving the $f_{2}$ are opposite. If there is no mixing in the $B^{0} \rightarrow f^{'}_{2}f_{0}[ \rightarrow \pi^ {+}\pi^{-}]$decay process, the $CP$ asymmetry is zero due to the pure penguin contribution from the annihilation diagram. However, when both the $\phi$ and $\theta$ mixing angles are taken into consideration simultaneously, they provide $f^{n}_{2}$ and $f^{n}$ components, leading to a significant interference between the tree operator contribution and the penguin contribution. This interference results in a large $CP$ asymmetry and the branching ratio increases by $1$ order of magnitude. The $\phi$ is really small, so the decay branching ratios involving $f_{2}$ barely change. The branching ratio of $B^{0} \rightarrow f_{2}f_{0}[ \rightarrow \pi^ {+}\pi^{-}]$ is bigger than that of $B^{0} \rightarrow f^{'}_{2}f_{0}[ \rightarrow \pi^ {+}\pi^{-}]$ with the reason that the $\frac{1}{\sqrt{2}}(u\overline{u}+d\overline{d})$ component makes the dominant contribution in the $B^{0}$ decays.

In Ref.~\cite{L:2021vb}, the authors have taken the mixing angle $\theta=145^{\circ}$, studied the branching fractions of the $B^{0} \rightarrow Vf_{0}[ \rightarrow \pi^{+}\pi^{-}$ decays in the pQCD approach, and gotten the results as follows:

\begin{equation}
\begin{split}
{\cal B}(B^{0} \rightarrow \rho^{0}f_{0}[ \rightarrow \pi^ {+}\pi^{-}])=0.82^{+0.36+0.02+0.05}_{-0.34-0.16-0.10}\times10^{-6} \\
{\cal B}(B^{0} \rightarrow \omega f_{0}[ \rightarrow \pi^ {+}\pi^{-}])=0.97^{+0.51+0.16+0.13}_{-0.39-0.19-0.10}\times10^{-6}\\
{\cal B}(B^{0} \rightarrow  K^{*0}f_{0}[ \rightarrow \pi^ {+}\pi^{-}])=0.82^{+0.36+0.02+0.05}_{-0.34-0.16-0.10}\times10^{-6} \\
\end{split}
\end{equation}
These are comparable to our results of $B^{0} \rightarrow (a^{0}_{2},f_{2},K^{*}_{2})f_{0}[ \rightarrow\pi^ {+}\pi^{-}]$ decays which are shown in Table \ref{f} because $\rho$ and $a^{0}_{2}$, $\omega$ and $f_{2}$, and $K^{*0}$ and $K^{*}_{2}$ have the same components in the quark model. The branching ratio of $B^{0} \rightarrow (\rho^{0},\omega,K^{*0})f_{0}[ \rightarrow \pi^ {+}\pi^{-}]$ and $B^{0} \rightarrow (a^{0}_{2},f_{2},K^{*}_{2})f_{0}[ \rightarrow\pi^ {+}\pi^{-}]$ are at the same order but the former is a little small. The reason is that the QCD dynamics of the vector mesons and tensor mesons are different, the tensor meson has a greater mass than vector meson, and we take the smaller slightly mixing angle $\theta=135^{\circ}$. As shown in Fig.\ref{fig4}, when the mixing angle $\theta > 90^{\circ}$, the contribution from the $\frac{1}{\sqrt{2}}(u\bar{u} + d\bar{d})$ component can cancel the contributions from the $s\bar{s}$ component, and the branching ratio of $B^{0} \rightarrow Tf_{0}[ \rightarrow\pi^ {+}\pi^{-}]$ decreases as the angle increases. We expect these results can be tested by the LHCb and Belle II experiments in the near future.

\begin{figure}[htbp]
	\centering
	\begin{tabular}{l}
		\includegraphics[width=0.5\textwidth]{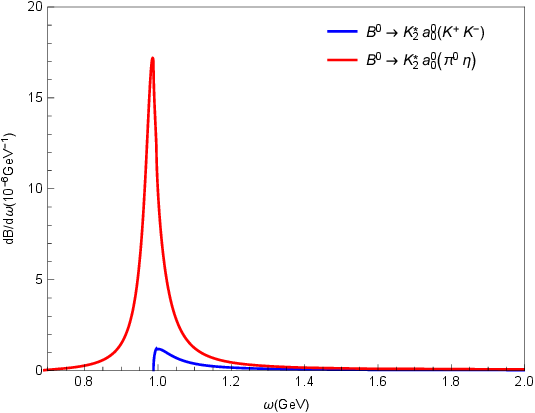}
		\includegraphics[width=0.5\textwidth]{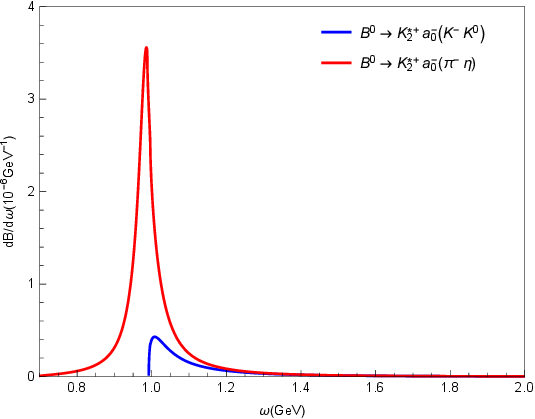}
	\end{tabular}
	\caption {Differential branching fractions of the ${B}^{0}\rightarrow K^{*}_{2}a_{0}[ \rightarrow K\overline{K}, \pi\eta]$ decays.}
	\label{fig6}
\end{figure}

 \begin{figure}[htbp]
	\centering
	\begin{tabular}{l}
		\includegraphics[width=0.5\textwidth]{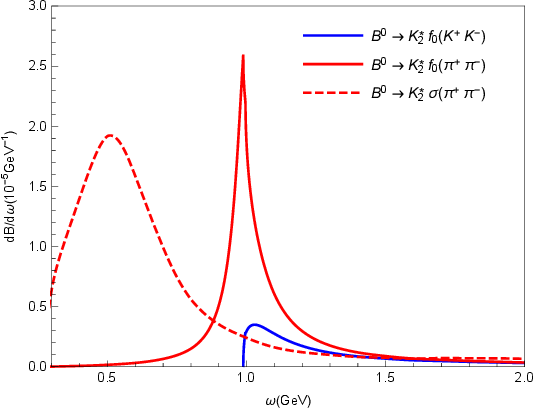}
	\end{tabular}
	\caption {Differential branching fractions of the ${B}^{0} \rightarrow K^{*}_{2}f_{0}[ \rightarrow \pi^{+}\pi^{-},K^{+}K^{-}]$ and ${B}^{0} \rightarrow K^{*}_{2}\sigma[ \rightarrow \pi^{+}\pi^{-}]$ decays.}
	\label{fig7}
\end{figure}

Different from the fixed kinematics of the two-body decays, the decay amplitudes of the quasi-two-body decays depend on the invariant mass, which results in the differential distribution of branching ratios. In Fig.~\ref{fig6}, we graph the differential branching ratios of the ${B}^{0}\rightarrow K^{*}_{2}a_{0}[ \rightarrow  K\overline{K}, \pi\eta]$ decays on the invariant mass; the results of the ${B}^{0}\rightarrow K^{*}_{2}a^{0}_{0}[ \rightarrow K^{+}K^{-}, \pi^{0}\eta]$ models are shown on the left, and those of the ${B}^{0} \rightarrow K^{*+}_{2}a^{-}_{0}[ \rightarrow K^{-}K^{0}, \pi^{-}\eta]$ models are shown on the right. The differential branching ratios of the  ${B}^{0} \rightarrow K^{*}_{2}f_{0}[ \rightarrow \pi^{+}\pi^{-}]$, ${B}^{0} \rightarrow K^{*}_{2}f_{0}[ \rightarrow K^{+}K^{-}]$, and ${B}^{0} \rightarrow K^{*}_{2}\sigma[ \rightarrow \pi^{+}\pi^{-}]$ decays on the $\pi\pi$ or $K\overline{K}$ invariant mass $\omega$ are presented in Fig.~\ref{fig7} with a red solid line, blue solid line, or red dashed line, respectively. For the $a_{0}$ resonance, the contributions of the $K\overline{K}$ channel are much smaller than that of the $\pi\eta$ channel. From the figures, it is clear that the peak occurs around the resonance peak mass; the majority of the branching ratios are concentrated around the resonance state, basically in the range of $[m_{S}-\Gamma_{S}, m_{S}+\Gamma_{S}]$; and the contributions from the $\omega\textgreater 2.0GeV$ are really small and can be neglected safely.

\section{Summary} \label{sec:summary}

In this article, considering the mixing angle $\theta=135^\circ$ between $f_{0}(980)$ and $f_{0}(500)$, and $\phi=9^\circ$ between $f_{2}(1270) $and $f^{'}_{2}(1525)$, we predict the branching fractions and direct $CP$ asymmetries of the $B^{0}\rightarrow Ta_{0}[ \rightarrow  K\overline{K}, \pi\eta]$, ${B}^{0} \rightarrow Tf_{0}[ \rightarrow \pi^{+}\pi^{-}, K^{+}K^{-}]$ and ${B}^{0} \rightarrow T\sigma[ \rightarrow \pi^{+}\pi^{-}]$ decays with the pQCD approach firstly, where $T$ denotes tensor mesons $a_{2}(1320)$, $K^{*}_{2}(1430)$, $f_{2}(1270)$, and $f^{'}_{2}(1525)$; the scalars are considered as the $q\overline{q}$ state in the first scenario. Our results show that: among these considered decays, the pQCD predictions for the $CP$-averaged branching ratios vary in the range of $10^{-8}$ to $10^{-5}$. The branching fractions are sensitive to the $\theta$, and the opposite is true for $\phi$, except for some decays involving $f^{'}_{2}(1525)$. The direct $CP$ asymmetries are also affected by the mixing angle. Using the narrow-width approximation, we calculate the relative partial decay widths ${\Gamma}(a_{0} \rightarrow K\overline{K})/{\Gamma}(a_{0} \rightarrow \pi\eta)$ and the ratio ${\cal B}(f_{0} \rightarrow K^{+}K^{-})/{\cal B}(f_{0} \rightarrow \pi^{+}\pi^{-})$, and extract the decays $B^{0}\rightarrow K^{*}_{2}(1430)^{0}f_{0}(980)$ branching fractions, which are in agreement with the existing experimental and theoretical values. Our work has positive implications for understanding the QCD behavior of the scalars, and we also expect that our calculations can be tested by the LHCb and Belle II experiments in the future.


\section*{acknowledgments}

This work is supported by the National Natural Science Foundation of China under Grant No.11047028.
\section*{APPENDIX : FACTORIZATION FORMULAS} \label{sec:appendix}
\appendix
\setcounter{equation}{0}
\renewcommand\theequation{A.\arabic{equation}}

In this section, we list the factorization formulas with $\overline{x}_{i}=1-x_{i}, \overline{\eta}=1-\eta$. $C_{F}=\frac{4}{3}$ stands for the color factor. Because of the charge conjugation invariance, the $S$-wave meson pair cannot be emitted with the $(V-A)(V-A)$ and $(V-A)(V+A)$ current, namely

\begin{eqnarray}
F^{LL}_{S}=F^{LR}_{S}=0
\end{eqnarray}

\begin{eqnarray}
\begin{split}
F^{SP}_{S}=&16\sqrt{\frac{2}{3}}\pi C_{F}\overline{f}_{s}M^{4}_{B}\int_{0}^{1}dx_{B}dx_{3}\int_{0}^{\infty}b_{B}db_{B}b_{3}db_{3}\phi_{B}(x_{B},b_{B})\\
&\times\{ [rx_{3}\overline{\eta}(r^{2}-\overline{\eta})\phi^{t}_{T}+r(4r_{b}-2+x_{3}(r^{2}+\overline{\eta}))\phi^{s}_{T}+(\overline{\eta}-r^{2})(2-r_{b})\phi_{T}]\\
&\times \alpha_{s}(t_{a1})h_{a1}(\alpha_{a1}, \beta_{a1}, b_{B}, b_{3})\exp[-S_{B}(t_{a1})-S_{3}(t_{a1})]S_{t}(x_{3})\\
&+[2r(\overline{\eta}-x_{B}+r^{2})\phi^{s}_{T}-r^{2}x_{B}\phi_{T}]\alpha_{s}(t_{b1})h_{b1}(\alpha_{b1},\beta_{b1}, b_{3}, b_{B})\exp[-S_{B}(t_{b1})-S_{3}(t_{b1})]S_{t}(x_{B})\}.
\end{split}
\end{eqnarray}

\begin{eqnarray}
\begin{split}
M^{LL}_{S}=&-\frac{32}{3}\pi C_{F}M^{4}_{B}\int_{0}^{1}dx_{B}dzdx_{3}\int_{0}^{\infty}b_{B}db_{B}bdb\phi_{B}(x_{B},b_{B})\phi_{S}\\
&\times\{[r((r^{2}-1)(x_{B}+(z-2)\eta)+x_{3}\overline{\eta}+r^{2}x_{3}(2\eta-1))\phi^{s}_{T}+r\overline{\eta}((1-r^{2})(x_{B}-z\eta)+x_{3}(r^{2}-\overline{\eta}))\phi^{t}_{T}\\
&+(r^{2}-1-\eta)(\bar{z}\bar{\eta}+r^{2}(z\overline{\eta}-\overline{x}_{B}))\phi_{T}]\alpha_{s}(t_{c1})h_{c1}(\alpha_{c1}, \beta_{c1}, b, b_{B})\exp[-S_{B}(t_{c1})-S(t_{c1})-S_{3}(t_{c1})]\\
&+[r((r^{2}-1)(x_{B}-z\eta)+x_{3}\overline{\eta}+r^{2}x_{3}(2\eta-1))\phi^{s}_{T}+r\overline{\eta}((r^{2}-1)(x_{B}+z\eta)-x_{3}(r^{2}-\overline{\eta}))\phi^{t}_{T}\\
&+(r^{2}-1+\eta)((1-r^{2})z-x_{B}+x_{3}(r^{2}+\overline{\eta}))\phi_{T}]\alpha_{s}(t_{d1})h_{d1}(\alpha_{d1}, \beta_{d1}, b, b_{B})\exp[-S_{B}(t_{d1})-S(t_{d1})-S_{3}(t_{d1})]\}
\end{split}
\end{eqnarray}

\begin{eqnarray}
\begin{split}
M^{LR}_{S}=&-\frac{32}{3}\pi C_{F}M^{4}_{B}\int_{0}^{1}dx_{B}dzdx_{3}\int_{0}^{\infty}b_{B}db_{B}bdb\phi_{B}(x_{B},b_{B})\\
&\times\{[2(\overline{z}+r^{2}(x_{3}-\overline{z}))(\overline{\eta}\phi_{T}+r\phi^{s}_{T}-r\overline{\eta}\phi^{t}_{T})\phi^{t}_{S}+((\bar{z}\bar{\eta}+r^{2}(z\overline{\eta}-\overline{x}_{B}))\phi_{T}\\
&+r(\overline{x}_{B}-z+x_{3}+r^{2}(x_{3}-\overline{z})+\overline{x}_{3}\eta)\phi^{s}_{T}-r\overline{\eta}(\overline{x}_{3}\overline{\eta}+x_{B}-z+r^{2}(x_{3}-\overline{z}))\phi^{t}_{T})(\phi^{s}_{S}-\phi^{t}_{S})]\\
&\times\alpha_{s}(t_{c1})h_{c1}(\alpha_{c1}, \beta_{c1}, b, b_{B})\exp[-S_{B}(t_{c1})-S(t_{c1})-S_{3}(t_{c1})]-[2r(x_{B}-x_{3}\overline{\eta})(r\phi_{T}-\phi^{s}_{T}-\overline{\eta}\phi^{t}_{T})\phi^{t}_{S}\\
&+((z\overline{\eta}+r^{2}(x_{B}-z\overline{\eta}))\phi_{T}+r(z(1-r^{2})-x_{B}+x_{3}(\overline{\eta}+r^{2}))\phi^{s}_{T}-r\overline{\eta}(z(1-r^{2})+x_{B}+x_{3}(r^{2}-\overline{\eta}))\phi^{t}_{T})\\
&\times(\phi^{s}_{S}-\phi^{t}_{S})]\alpha_{s}(t_{d1})h_{d1}(\alpha_{d1}, \beta_{d1}, b, b_{B})\exp[-S_{B}(t_{d1})-S(t_{d1})-S_{3}(t_{d1})]\}
\end{split}
\end{eqnarray}

\begin{eqnarray}
\begin{split}
M^{SP}_{S}=&-\frac{32}{3}\pi C_{F}M^{4}_{B}\int_{0}^{1}dx_{B}dzdx_{3}\int_{0}^{\infty}b_{B}db_{B}bdb\phi_{B}(x_{B},b_{B})\phi_{S}\\
&\times\{[r((r^{2}-1)(x_{B}+(z-2)\eta)+x_{3}\overline{\eta}+r^{2}x_{3}(2\eta-1))\phi^{s}_{T}-r\overline{\eta}((1-r^{2})(x_{B}-z\eta)+x_{3}(r^{2}-\overline{\eta}))\phi^{t}_{T}\\
&+(r^{2}-\overline{\eta})(\overline{x}_{B}+x_{3}-z+r^{2}(x_{3}-\overline{z})+\eta\overline{x}_{3})\phi_{T}]\alpha_{s}(t_{c1})h_{c1}(\alpha_{c1}, \beta_{c1}, b, b_{B})\exp[-S_{B}(t_{c1})-S(t_{c1})-S_{3}(t_{c1})]\\
&-[r((r^{2}-1)(x_{B}-z\eta)+x_{3}\overline{\eta}+r^{2}x_{3}(2\eta-1))\phi^{s}_{T}-r\overline{\eta}((r^{2}-1)(x_{B}+z\eta)-x_{3}(r^{2}-\overline{\eta}))\phi^{t}_{T}\\
&+(r^{2}-1-\eta)(z\overline{\eta}+r^{2}(x_{B}-z\overline{\eta}))\phi_{T}]\alpha_{s}(t_{d1})h_{d1}(\alpha_{d1}, \beta_{d1}, b, b_{B})\exp[-S_{B}(t_{d1})-S(t_{d1})-S_{3}(t_{d1})]\}
\end{split}
\end{eqnarray}

\begin{eqnarray}
\begin{split}
A^{LL}_{S}=&8\sqrt{\frac{2}{3}}\pi C_{F}f_{B}M^{4}_{B}\int_{0}^{1}dzdx_{3}\int_{0}^{\infty}bdbb_{3}db_{3}\\
&\times\{[2r\sqrt{(1-r^{2})\eta}x_{3}\bar{\eta}(\bar{\eta}-r^{2})\phi^{s}_{S}\phi^{t}_{T}-2r\sqrt{(1-r^{2})\eta}(x_{3}(r^{2}+\bar{\eta})-2)\phi^{s}_{S}\phi^{s}_{T}\\
&+(\bar{\eta}-r^{2})(x_{3}\bar{\eta}-1+r^{2}x_{3}\eta)\phi_{S}\phi_{T}]\alpha_{s}(t_{e1})h_{e1}(\alpha_{e1}, \beta_{e1}, b, b_{3})\exp[-S_{3}(t_{e1})-S(t_{e1})]S_{t}(x_{3})\\
&-[2r\sqrt{(1-r^{2})\eta}(r^{2}\bar{z}-\bar{z}+\eta)\phi^{t}_{S}\phi^{s}_{T}+2r\sqrt{(1-r^{2})\eta}(r^{2}\bar{z}+z+\bar{\eta})\phi^{s}_{S}\phi^{s}_{T}\\
&+\bar{\eta}(r^{4}\bar{z}-z+r^{2}(2z-\bar{\eta}))\phi_{S}\phi_{T}]\alpha_{s}(t_{f1})h_{f1}(\alpha_{f1}, \beta_{f1}, b_{3}, b)\exp[-S_{3}(t_{f1})-S(t_{f1})]S_{t}(z)\}
\end{split}
\end{eqnarray}

\begin{eqnarray}
A^{LR}_{S}=A^{LL}_{S}
\end{eqnarray}

\begin{eqnarray}
\begin{split}
A^{SP}_{S}=&-16\sqrt{\frac{2}{3}}\pi C_{F}f_{B}M^{4}_{B}\int_{0}^{1}dzdx_{3}\int_{0}^{\infty}bdbb_{3}db_{3}\\
&\times \{[r(x_{3}\bar{\eta}-1-\eta+r^{2}(\bar{x}_{3}+2\eta x_{3}))\phi_{S}\phi^{s}_{T}-2\sqrt{(1-r^{2})\eta}(r^{2}-\bar{\eta})\phi^{s}_{S}\phi_{T}\\
&+r\bar{x}_{3}\bar{\eta}(r^{2}-\bar{\eta})\phi_{S}\phi^{t}_{T}]\alpha_{s}(t_{e1})h_{e1}(\alpha_{e1}, \beta_{e1}, b, b_{3})\exp[-S_{3}(t_{e1})-S(t_{e1})]S_{t}(x_{3})\\
&-[2r^{2}\sqrt{(1-r^{2})\eta}\bar{\eta}\phi^{t}_{S}\phi_{T}-2r(\bar{z}\eta-1+r^{2}(1-2\eta+z\eta))\phi_{S}\phi^{s}_{T}\\
&+\sqrt{(1-r^{2})\eta}(r^{2}-1)z\bar{\eta}(\phi^{s}_{S}-\phi^{t}_{S})\phi_{T}]\alpha_{s}(t_{f1})h_{f1}(\alpha_{f1}, \beta_{f1}, b_{3}, b)\exp[-S_{3}(t_{f1})-S(t_{f1})]S_{t}(z)\}
\end{split}
\end{eqnarray}

\begin{eqnarray}
\begin{split}
W^{LL}_{S}=&-\frac{32}{3}\pi C_{F}M^{4}_{B}\int_{0}^{1}dx_{B}dzdx_{3}\int_{0}^{\infty}b_{B}db_{B}bdb\phi_{B}(x_{B},b_{B})\\
&\times\{[(r_{b}(\bar{\eta}-r^{2})+(1-r^{2}+\eta)(\eta-\bar{z}+r^{2}(\bar{z}-x_{B})+(r^{2}-1)z\eta))\phi_{S}\phi_{T}+2r\sqrt{(1-r^{2})\eta}(\bar{z}+r^{2}(x_{3}-\bar{z}))\\
&\times\phi^{t}_{S}(\phi^{s}_{T}-\bar{\eta}\phi^{t}_{T})+r\sqrt{(1-r^{2})\eta}(\bar{x}_{B}+x_{3}-z+r^{2}(x_{3}-\bar{z})+\bar{x}_{3}\eta)(\phi^{s}_{S}-\phi^{t}_{S})\phi^{s}_{T}-4rr_{b}\sqrt{(1-r^{2})\eta}\phi^{s}_{S}\phi^{s}_{T}\\
&-r\bar{\eta}\sqrt{(1-r^{2})\eta}(x_{B}-z+r^{2}({x}_{3}-\bar{z})+\bar{x}_{3}\bar{\eta})(\phi^{s}_{S}-\phi^{t}_{S})\phi^{t}_{T}]\alpha_{s}(t_{g1})h_{g1}(\alpha_{g1}, \beta_{g1}, b_{B}, b)\\
&\times\exp[-S_{B}(t_{g1})-S(t_{g1})-S_{3}(t_{g1})]-[2r\sqrt{(1-r^{2})\eta}(x_{B}-\bar{x}_{3}\bar{\eta})\phi^{t}_{S}(\phi^{s}_{T}-\bar{\eta}\phi^{t}_{T})\\
&+(r^{2}-\bar{\eta})(x_{B}-\bar{x}_{3}\bar{\eta}-z\eta-r^{2}(x_{B}-\bar{x}_{3}(1-2\eta)-z\eta))\phi_{S}\phi_{T}\\
&+r\sqrt{(1-r^{2})\eta}(x_{B}-\bar{x}_{3}\bar{\eta}+r^{2}(x_{3}-\bar{z})-z)(\phi^{s}_{S}-\phi^{t}_{S})\phi^{s}_{T}+r\bar{\eta}\sqrt{(1-r^{2})\eta}(\bar{x}_{3}\bar{\eta}-x_{B}+r^{2}(x_{3}-\bar{z})-z)\\
&\times(\phi^{s}_{S}-\phi^{t}_{S})\phi^{t}_{T}]\alpha_{s}(t_{h1})h_{h1}(\alpha_{h1}, \beta_{h1}, b_{B}, b)\exp[-S_{B}(t_{h1})-S(t_{h1})-S_{3}(t_{h1})]\}
\end{split}
\end{eqnarray}

\begin{eqnarray}
\begin{split}
W^{LR}_{S}=&-\frac{32}{3}\pi C_{F}M^{4}_{B}\int_{0}^{1}dx_{B}dzdx_{3}\int_{0}^{\infty}b_{B}db_{B}bdb\phi_{B}(x_{B},b_{B})\\
&\times\{[2\bar{\eta}\sqrt{(1-r^{2})\eta}(\bar{z}+r^{2}(x_{3}-\bar{z})+r_{b})\phi^{t}_{S}\phi_{T}+r((1-r^{2})(r_{b}-x_{B}+x_{3})+(2+r_{b}-z-x_{3}+r^{2}(z-2\bar{x}_{3}))\eta)\\
&\times\phi_{S}\phi^{s}_{T}-\sqrt{(1-r^{2})\eta}(r^{2}(\bar{x}_{B}+r_{b}-z\bar{\eta})-(r_{b}+\bar{z})\bar{\eta})(\phi^{s}_{S}-\phi^{t}_{S})\phi_{T}+r\bar{\eta}((1-r^{2})(x_{B}-z\eta)+(x_{3}+r_{b})(r^{2}-\bar{\eta}))\\
&\times\phi_{S}\phi^{t}_{T}]\alpha_{s}(t_{g1})h_{g1}(\alpha_{g1}, \beta_{g1}, b_{B}, b)\exp[-S_{B}(t_{g1})-S(t_{g1})-S_{3}(t_{g1})]\\
&-[2\bar{\eta}\sqrt{(1-r^{2})\eta}(r^{2}(x_{3}-\bar{z})-z)\phi^{t}_{S}\phi_{T}-r(\bar{x}_{3}\bar{\eta}-x_{B}+z\eta+r^{2}(x_{B}-(1-2\eta)\bar{x}_{3}-z\eta))\phi_{S}\phi^{s}_{T}\\
&-\sqrt{(1-r^{2})\eta}(z\bar{\eta}+r^{2}(x_{B}-z\bar{\eta}))(\phi^{s}_{S}-\phi^{t}_{S})\phi_{T}+r
\bar{\eta}(\bar{x}_{3}\bar{\eta}-x_{B}+r^{2}(x_{3}-\bar{x}_{B}+z\eta)-z\eta)\phi_{S}\phi^{t}_{T}]\\
&\times\alpha_{s}(t_{h1})h_{h1}(\alpha_{h1}, \beta_{h1}, b_{B}, b)\exp[-S_{B}(t_{h1})-S(t_{h1})-S_{3}(t_{h1})]\}
\end{split}
\end{eqnarray}

\begin{eqnarray}
\begin{split}
W^{SP}_{S}=&\frac{32}{3}\pi C_{F}M^{4}_{B}\int_{0}^{1}dx_{B}dzdx_{3}\int_{0}^{\infty}b_{B}db_{B}bdb\phi_{B}(x_{B},b_{B})\\
&\times\{[(r^{2}-\bar{\eta})(r_{b}+(1-r^{2})(x_{B}-x_{3})+(z-2+x_{3}-r^{2}(z-2\bar{x}_{3}))\eta)\phi_{S}\phi_{T}+2r\sqrt{(1-r^{2})\eta}(x_{B}-x_{3}\bar{\eta}-\eta)\\
&\times\phi^{t}_{S}(\phi^{s}_{T}-\bar{\eta}\phi^{t}_{T})+4rr_{b}\sqrt{(1-r^{2})\eta}\phi^{s}_{S}\phi^{s}_{T}-r\bar{\eta}\sqrt{(1-r^{2})\eta}(x_{B}+\bar{x}_{3}\bar{\eta}-z+r^{2}(x_{3}-\bar{z}))(\phi^{s}_{S}-\phi^{t}_{S})\phi^{t}_{T}\\
&-r\sqrt{(1-r^{2})\eta}(\bar{x}_{B}+x_{3}-z+r^{2}(x_{3}-\bar{z})+\bar{x}_{3}\eta)(\phi^{s}_{S}-\phi^{t}_{S})\phi^{s}_{T}]\alpha_{s}(t_{g1})h_{g1}(\alpha_{g1}, \beta_{g1}, b_{B}, b)\exp[-S_{B}(t_{g1})\\
&-S(t_{g1})-S_{3}(t_{g1})]-[r\bar{\eta}\sqrt{(1-r^{2})\eta}(\bar{x}_{3}\bar{\eta}-x_{B}-z+r^{2}(x_{3}-\bar{z}))(\phi^{s}_{S}-\phi^{t}_{S})\phi^{t}_{T}-2r\sqrt{(1-r^{2})\eta}(r^{2}(x_{3}-\bar{z})\\
&-z)\phi^{t}_{S}(\phi^{s}_{T}-\bar{\eta}\phi^{t}_{T})+(r^{2}-1-\eta)(z\bar{\eta}+r^{2}(x_{B}-z\bar{\eta}))\phi_{S}\phi_{T}-r\sqrt{(1-r^{2})\eta}(x_{B}-\bar{x}_{3}\bar{\eta}-z+r^{2}(x_{3}-\bar{z}))\\
&\times(\phi^{s}_{S}-\phi^{t}_{S})\phi^{s}_{T}]\alpha_{s}(t_{h1})h_{h1}(\alpha_{h1}, \beta_{h1}, b_{B}, b)\exp[-S_{B}(t_{h1})-S(t_{h1})-S_{3}(t_{h1})]\}
\end{split}
\end{eqnarray}

\begin{eqnarray}
\begin{split}
M^{LL}_{T}=&-\frac{32}{3}\pi C_{F}M^{4}_{B}\int_{0}^{1}dx_{B}dzdx_{3}\int_{0}^{\infty}b_{B}db_{B}b_{3}db_{3}\phi_{B}(x_{B},b_{B})\phi_{T}\\
&\times\{[2r^{2}\sqrt{(1-r^{2})\eta}(x_{B}-\bar{x}_{3}\bar{\eta})\phi^{t}_{S}+\sqrt{(1-r^{2})\eta}(z\bar{\eta}+r^{2}(x_{B}-z\bar{\eta}))(\phi^{s}_{S}-\phi^{t}_{S})+(\bar{\eta}-r^{2})\\
&\times(x_{B}-\bar{x}_{3}\bar{\eta}-z\eta-r^{2}(x_{B}-\bar{x}_{3}+2\eta\bar{x}_{3}-z\eta))\phi_{S}]\alpha_{s}(t_{c2})h_{c2}(\alpha_{c2}, \beta_{c2}, b_{3}, b_{B})\exp[-S_{B}(t_{c2})-S(t_{c2})\\
&-S_{3}(t_{c2})]-[\sqrt{(1-r^{2})\eta}(z\bar{\eta}+r^{2}(x_{B}-z\bar{\eta}))(\phi^{s}_{S}-\phi^{t}_{S})-2\bar{\eta}\sqrt{(1-r^{2})\eta}(z(r^{2}-1)-r^{2}x_{3})\phi^{t}_{S}+(\bar{\eta}-r^{2})\\
&\times(x_{B}+z(r^{2}-1)-(\bar{\eta}+r^{2})x_{3})\phi_{S}]\alpha_{s}(t_{d2})h_{d2}(\alpha_{d2}, \beta_{d2}, b_{3}, b_{B})\exp[-S_{B}(t_{d2})-S(t_{d2})-S_{3}(t_{d2})]\}
\end{split}
\end{eqnarray}

\begin{eqnarray}
\begin{split}
M^{LR}_{T}=&-\frac{32}{3}\pi C_{F}rM^{4}_{B}\int_{0}^{1}dx_{B}dzdx_{3}\int_{0}^{\infty}b_{B}db_{B}b_{3}db_{3}\phi_{B}(x_{B},b_{B})\\
&\times\{[2\sqrt{(1-r^{2})\eta}(z-r^{2}(x_{3}-\bar{z}))\phi^{t}_{S}(\phi^{s}_{T}-\bar{\eta}\phi^{t}_{T})+(\bar{x}_{3}\bar{\eta}-x_{B}+z\eta+r^{2}(x_{B}-\bar{x}_{3}+2\eta\bar{x}_{3}-z\eta))\phi_{S}\phi^{s}_{T}\\
&-\sqrt{(1-r^{2})\eta}(x_{B}-\bar{x}_{3}\bar{\eta}-z+r^{2}(x_{3}-\bar{z}))(\phi^{s}_{S}-\phi^{t}_{S})\phi^{s}_{T}+\bar{\eta}(\bar{x}_{3}\bar{\eta}-x_{B}-z\eta+r^{2}(x_{B}-\bar{x}_{3}+z\eta))\phi_{S}\phi^{t}_{T}\\
&+\bar{\eta}\sqrt{(1-r^{2})\eta}(\bar{x}_{3}\bar{\eta}-x_{B}-z+r^{2}(x_{3}-\bar{z}))(\phi^{s}_{S}-\phi^{t}_{S})\phi^{t}_{T}]\alpha_{s}(t_{c2})h_{c2}(\alpha_{c2}, \beta_{c2}, b_{3}, b_{B})\exp[-S_{B}(t_{c2})\\
&-S(t_{c2})-S_{3}(t_{c2})]-[((r^{2}-1)(x_{B}-z\eta)+x_{3}\bar{\eta}+r^{2}x_{3}(2\eta-1))\phi_{S}\phi^{s}_{T}-2\sqrt{(1-r^{2})\eta}(z(r^{2}-1)-r^{2}x_{3})\\
&\times\phi^{t}_{S}(\phi^{s}_{T}+\bar{\eta}\phi^{t}_{T})-\bar{\eta}((r^{2}-1)(x_{B}+z\eta-x_{3})-x_{3}\eta)\phi_{S}\phi^{t}_{T}-\sqrt{(1-r^{2})\eta}(x_{B}+z(r^{2}-1)-(r^{2}+\bar{\eta})x_{3})\\
&\times(\phi^{s}_{S}-\phi^{t}_{S})\phi^{s}_{T}+\bar{\eta}\sqrt{(1-r^{2})\eta}(x_{B}+(1-r^{2})(z-x_{3})+x_{3}\eta)(\phi^{s}_{S}-\phi^{t}_{S})\phi^{t}_{T}]\\
&\times\alpha_{s}(t_{d2})h_{d2}(\alpha_{d2}, \beta_{d2}, b_{3}, b_{B})\exp[-S_{B}(t_{d2})-S(t_{d2})-S_{3}(t_{d2})]\}
\end{split}
\end{eqnarray}

\begin{eqnarray}
\begin{split}
M^{SP}_{T}=&-\frac{32}{3}\pi C_{F}M^{4}_{B}\int_{0}^{1}dx_{B}dzdx_{3}\int_{0}^{\infty}b_{B}db_{B}b_{3}db_{3}\phi_{B}(x_{B},b_{B})\phi_{T}\\
&\times\{[\sqrt{(1-r^{2})\eta}(z\bar{\eta}+r^{2}(x_{B}-z\bar{\eta}))(\phi^{s}_{S}-\phi^{t}_{S})-2\bar{\eta}\sqrt{(1-r^{2})\eta}(r^{2}(x_{3}-\bar{z})-z)\phi^{t}_{S}\\
&+(\bar{\eta}-r^{2})(x_{B}-\bar{x}_{3}\bar{\eta}-z+r^{2}(x_{3}-\bar{z}))\phi_{S}]\alpha_{s}(t_{c2})h_{c2}(\alpha_{c2}, \beta_{c2}, b_{3}, b_{B})\exp[-S_{B}(t_{c2})-S(t_{c2})-S_{3}(t_{c2})]\\
&-[2r^{2}\sqrt{(1-r^{2})\eta}(x_{B}-x_{3}\bar{\eta})\phi^{t}_{S}+\sqrt{(1-r^{2})\eta}(z\bar{\eta}+r^{2}(x_{B}-z\bar{\eta}))(\phi^{s}_{S}-\phi^{t}_{S})+(r^{2}-\bar{\eta})\\
&\times((r^{2}-1)(x_{B}-z\eta)+x_{3}\bar{\eta}+r^{2}x_{3}(2\eta-1))\phi_{S}]\alpha_{s}(t_{d2})h_{d2}(\alpha_{d2}, \beta_{d2}, b_{3}, b_{B})\exp[-S_{B}(t_{d2})-S(t_{d2})-S_{3}(t_{d2})]\}
\end{split}
\end{eqnarray}

\begin{eqnarray}
\begin{split}
A^{LL}_{T}=&8\sqrt{\frac{2}{3}}\pi C_{F}f_{B}M^{4}_{B}\int_{0}^{1}dzdx_{3}\int_{0}^{\infty}bdbb_{3}db_{3}\\
&\times\{[2r\sqrt{(1-r^{2})\eta}(2+(r^{2}-1)z)(\phi^{s}_{S}-\phi^{t}_{S})\phi^{s}_{T}-(\bar{z}\bar{\eta}+r^{2}(2z\bar{\eta}-1)-r^{4}z\bar{\eta})\phi_{S}\phi_{T}\\
&+4r\sqrt{(1-r^{2})\eta}\phi^{t}_{S}\phi^{s}_{T}]\alpha_{s}(t_{e2})h_{e2}(\alpha_{e2}, \beta_{e2}, b_{3}, b)\exp[-S_{3}(t_{e2})-S(t_{e2})]S_{t}(z)\\
&-[(r^{2}-\bar{\eta})(x_{3}+(1-r^{2})\bar{x}_{3}\eta)\phi_{S}\phi_{T}+2r\sqrt{(1-r^{2})\eta}(1+x_{3}-r^{2}\bar{x}_{3}+\bar{x}_{3}\eta)\phi^{s}_{S}\phi^{s}_{T}\\
&+2r\sqrt{(1-r^{2})\eta}\bar{x}_{3}\bar{\eta}(r^{2}-\bar{\eta})\phi^{s}_{S}\phi^{t}_{T}]\alpha_{s}(t_{f2})h_{f2}(\alpha_{f2}, \beta_{f2}, b, b_{3})\exp[-S_{3}(t_{f2})-S(t_{f2})]S_{t}(x_{3})\}
\end{split}
\end{eqnarray}

\begin{eqnarray}
A^{LR}_{T}=A^{LL}_{T}
\end{eqnarray}

\begin{eqnarray}
\begin{split}
A^{SP}_{T}=&-16\sqrt{\frac{2}{3}}\pi C_{F}f_{B}M^{4}_{B}\int_{0}^{1}dzdx_{3}\int_{0}^{\infty}bdbb_{3}db_{3}\\
&\times\{[\sqrt{(1-r^{2})\eta}(r^{2}(1-z\bar{\eta})-\bar{\eta}\bar{z})(\phi^{s}_{S}-\phi^{t}_{S})\phi_{T}-2\bar{\eta}\sqrt{(1-r^{2})\eta}(1+(r^{2}-1)z)\phi^{t}_{S}\phi_{T}\\
&+2r(1+\bar{z}\eta+r^{2}(z\eta-1))\phi_{S}\phi^{s}_{T}]\alpha_{s}(t_{e2})h_{e2}(\alpha_{e2}, \beta_{e2}, b_{3}, b)\exp[-S_{3}(t_{e2})-S(t_{e2})]S_{t}(z)\\
&+[2\sqrt{(1-r^{2})\eta}(r^{2}-\bar{\eta})\phi^{s}_{S}\phi_{T}+r(x_{3}\bar{\eta}-2(r^{2}-1)\eta+r^{2}x_{3}(2\eta-1))\phi_{S}\phi^{s}_{T}\\
&+rx_{3}\bar{\eta}(r^{2}-\bar{\eta})\phi_{S}\phi^{t}_{T}]\alpha_{s}(t_{f2})h_{f2}(\alpha_{f2}, \beta_{f2}, b, b_{3})\exp[-S_{3}(t_{f2})-S(t_{f2})]S_{t}(x_{3})\}
\end{split}
\end{eqnarray}

\begin{eqnarray}
\begin{split}
W^{LL}_{T}=&-\frac{32}{3}\pi C_{F}M^{4}_{B}\int_{0}^{1}dx_{B}dzdx_{3}\int_{0}^{\infty}b_{B}db_{B}b_{3}db_{3}\phi_{B}(x_{B},b_{B})\\
&\times\{[r\sqrt{(1-r^{2})\eta}(\bar{\eta}(\bar{x}_{3}\bar{\eta}-x_{B}-z+r^{2}(x_{3}-\bar{z}))(\phi^{s}_{S}-\phi^{t}_{S})\phi^{t}_{T}-(x_{B}-\bar{x}_{3}\bar{\eta}-z+r^{2}(x_{3}-\bar{z}))(\phi^{s}_{S}-\phi^{t}_{S})\phi^{s}_{T}\\
&-2(r^{2}(x_{3}-\bar{z})-z)\phi^{t}_{S}(\phi^{s}_{T}-\bar{\eta}\phi^{t}_{T}))-(r_{b}+x_{B}-\bar{x}_{3}\bar{\eta}-z\eta-r^{2}(x_{B}+(2\eta-1)\bar{x}_{3}-z\eta))\\
&\times(r^{2}-\bar{\eta})\phi_{S}\phi_{T}-4rr_{b}\sqrt{(1-r^{2})\eta}\phi^{s}_{S}\phi^{s}_{T}]\alpha_{s}(t_{g2})h_{g2}(\alpha_{g2}, \beta_{g2}, b_{B}, b_{3})\exp[-S_{B}(t_{g2})-S(t_{g2})-S_{3}(t_{g2})]\\
&-[2r\sqrt{(1-r^{2})\eta}(x_{B}-x_{3}\bar{\eta}-\eta)\phi^{t}_{S}(\phi^{s}_{T}-\bar{\eta}\phi^{t}_{T})-r\sqrt{(1-r^{2})\eta}(\bar{x}_{B}-z+x_{3}+r^{2}(x_{3}-\bar{z})+\eta\bar{x}_{3})(\phi^{s}_{S}-\phi^{t}_{S})\phi^{s}_{T}\\
&+(r^{2}-1-\eta)(r^{2}(\bar{x}_{B}-z\bar{\eta})-\bar{\eta}\bar{z})\phi_{S}\phi_{T}-r\bar{\eta}\sqrt{(1-r^{2})\eta}(\bar{x}_{3}\bar{\eta}+x_{B}-z+r^{2}(x_{3}-\bar{z}))(\phi^{s}_{S}-\phi^{t}_{S})\phi^{t}_{T}]\\
&\times\alpha_{s}(t_{h2})h_{h2}(\alpha_{h2}, \beta_{h2}, b_{B}, b_{3})\exp[-S_{B}(t_{h2})-S(t_{h2})-S_{3}(t_{h2})]\}
\end{split}
\end{eqnarray}

\begin{eqnarray}
\begin{split}
W^{LR}_{T}=&-\frac{32}{3}\pi C_{F}M^{4}_{B}\int_{0}^{1}dx_{B}dzdx_{3}\int_{0}^{\infty}b_{B}db_{B}b_{3}db_{3}\phi_{B}(x_{B},b_{B})\\
&\times\{[2r^{2}\sqrt{(1-r^{2})\eta}(x_{B}-\bar{x}_{3}\bar{\eta}-r_{b})\phi^{t}_{S}\phi_{T}+r((1-r^{2})(\bar{x}_{3}-x_{B}+r_{b})+(z-\bar{x}_{3}-r^{2}(z-2\bar{x}_{3})+r_{b})\eta)\phi_{S}\phi^{s}_{T}\\
&+r\bar{\eta}(\bar{x}_{3}\bar{\eta}-x_{B}-z\eta+r^{2}(x_{B}-\bar{x}_{3}+z\eta)-r_{b}(r^{2}-\bar{\eta}))\phi_{S}\phi^{t}_{T}+\sqrt{(1-r^{2})\eta}(z\bar{\eta}+r^{2}(x_{B}-z\bar{\eta})-r_{b}(r^{2}-\bar{\eta}))\\
&\times(\phi^{s}_{S}-\phi^{t}_{S})\phi_{T}]\alpha_{s}(t_{g2})h_{g2}(\alpha_{g2}, \beta_{g2}, b_{B}, b_{3})\exp[-S_{B}(t_{g2})-S(t_{g2})-S_{3}(t_{g2})]\\
&-[r\bar{\eta}((1-r^{2})(x_{B}-z\eta)+x_{3}(r^{2}-\bar{\eta}))\phi_{S}\phi^{t}_{T}-r((r^{2}-1)(x_{B}+(z-2)\eta)+x_{3}\bar{\eta}+(2\eta-1)r^{2}x_{3})\phi_{S}\phi^{s}_{T}\\
&-\sqrt{(1-r^{2})\eta}(\bar{z}\bar{\eta}+r^{2}(z\bar{\eta}-\bar{x}_{B}))(\phi^{s}_{S}-\phi^{t}_{S})\phi_{T}-2r^{2}\sqrt{(1-r^{2})\eta}(x_{B}-\eta-x_{3}\bar{\eta})\phi^{t}_{S}\phi_{T}]\\
&\times\alpha_{s}(t_{h2})h_{h2}(\alpha_{h2}, \beta_{h2}, b_{B}, b_{3})\exp[-S_{B}(t_{h2})-S(t_{h2})-S_{3}(t_{h2})]\}
\end{split}
\end{eqnarray}

\begin{eqnarray}
\begin{split}
W^{SP}_{T}=&\frac{32}{3}\pi C_{F}M^{4}_{B}\int_{0}^{1}dx_{B}dzdx_{3}\int_{0}^{\infty}b_{B}db_{B}b_{3}db_{3}\phi_{B}(x_{B},b_{B})\\
&\times[(r_{b}(r^{2}-\bar{\eta})+(1-r^{2}+\eta)(z\bar{\eta}+r^{2}(x_{B}-z\bar{\eta})))\phi_{S}\phi_{T}+2r\sqrt{(1-r^{2})\eta}(x_{B}-\bar{x}_{3}\bar{\eta})\phi^{t}_{S}(\phi^{s}_{T}-\bar{\eta}\phi^{t}_{T})\\
&+r\sqrt{(1-r^{2})\eta}(\phi^{s}_{S}-\phi^{t}_{S})((x_{B}-z-\bar{x}_{3}\bar{\eta}+r^{2}(x_{3}-\bar{z}))\phi^{s}_{T}+\bar{\eta}(\bar{x}_{3}\bar{\eta}-x_{B}-z+r^{2}(x_{3}-\bar{z}))\phi^{t}_{T})\\
&+4rr_{b}\sqrt{(1-r^{2})\eta}\phi^{s}_{S}\phi^{s}_{T}]\alpha_{s}(t_{g2})h_{g2}(\alpha_{g2}, \beta_{g2}, b_{B}, b_{3})\exp[-S_{B}(t_{g2})-S(t_{g2})-S_{3}(t_{g2})]\\
&-[2r\sqrt{(1-r^{2})\eta}(\bar{z}+r^{2}(x_{3}-\bar{z}))\phi^{t}_{S}\phi^{s}_{T}+(r^{2}-\bar{\eta})((r^{2}-1)(x_{B}+(z-2)\eta)+x_{3}\bar{\eta}+r^{2}x_{3}(2\eta-1))\phi_{S}\phi_{T}\\
&+r\sqrt{(1-r^{2})\eta}(\phi^{s}_{S}-\phi^{t}_{S})((\bar{z}-x_{B}+x_{3}+r^{2}(x_{3}-\bar{z})+\eta\bar{x}_{3})\phi^{s}_{T}-\bar{\eta}(x_{B}-z+\bar{x}_{3}\bar{\eta}+r^{2}(x_{3}-\bar{z}))\phi^{t}_{T})\\
&-2r\bar{\eta}\sqrt{(1-r^{2})\eta}(\bar{z}+r^{2}(x_{3}-\bar{z}))\phi^{t}_{S}\phi^{t}_{T}]\alpha_{s}(t_{h2})h_{h2}(\alpha_{h2}, \beta_{h2}, b_{B}, b_{3})\exp[-S_{B}(t_{h2})-S(t_{h2})-S_{3}(t_{h2})]
\end{split}
\end{eqnarray}

The hard functions $h_{i}$ are derived from the Fourier transform with $i=(a1, ... ,h2)$, whose specific expression is

\begin{eqnarray}
\begin{split}
h_{i}(\alpha, \beta, b_{1}, b_{2})&=h_{1}(\alpha, b_{1})\times h_{2}(\beta, b_{1}, b_{2}),\\
h_{1}(\alpha, b_{1})&=
\begin{cases}
K_{0}(\sqrt{\alpha}b_{1}),&\alpha\textgreater 0,\\
K_{0}(i\sqrt{-\alpha}b_{1}),&\alpha\textless 0,
\end{cases}\\
h_{2}(\beta, b_{1}, b_{2})&=
\begin{cases}
\theta(b_{1}-b_{2})I_{0}(\sqrt{\beta}b_{2})K_{0}(\sqrt{\beta}b_{1})+(b_{1}\leftrightarrow b_{2}),&\beta\textgreater 0,\\
\theta(b_{1}-b_{2})J_{0}(\sqrt{-\beta}b_{2})K_{0}(i\sqrt{-\beta}b_{1})+(b_{1}\leftrightarrow b_{2}),&\beta\textless 0,
\end{cases}
\end{split}
\end{eqnarray}
with the Bessel function $J_{0}$, and the modified Bessel functions $K_{0}$ and $I_{0}$. The expressions for $\alpha$ and $\beta$ in the hard functions are

\begin{eqnarray}
\begin{split}
\alpha_{(a1, b1)}=\beta_{(c1, d1)}&=M^{2}_{B}r^{2}x_{3}(x_{B}-\overline{\eta}x_{3}),\\
\beta_{a1}&=M^{2}_{B}[(r^{2}x_{3}-1)(1-x_{3}\bar{\eta})+r^{2}_{b}],\\
\beta_{b1}&=M^{2}_{B}r^{2}(x_{B}-\overline{\eta}),\\
\alpha_{c1}&=M^{2}_{B}[r^{2}(z-\overline{x}_{3})+\overline{z}](x_{B}-\eta-x_{3}\overline{\eta}),\\
\alpha_{d1}&=M^{2}_{B}[r^{2}(x_{3}-z)+z](x_{B}-x_{3}\overline{\eta}),\\
\alpha_{(e1, f1)}=\beta_{(g1, h1)}&=M^{2}_{B}\overline{x}_{3}\overline{\eta}[r^{2}(x_{3}-\overline{z})-z],\\
\beta_{e1}&=M^{2}_{B}(r^{2}x_{3}-1)(1-x_{3}\overline{\eta}),\\
\beta_{f1}&=M^{2}_{B}\overline{\eta}(-r^{2}\overline{z}-z),\\
\alpha_{g1}&=M^{2}_{B}\{[r^{2}(z-\overline{x}_{3})+\overline{z}](x_{B}-\eta-x_{3}\overline{\eta})+r^{2}_{b}\},\\
\alpha_{h1}&=M^{2}_{B}[r^{2}(\overline{x}_{3}-z)+z](x_{B}-\overline{x}_{3}\overline{\eta}),\\
\beta_{(c2, d2)}&=M^{2}_{B}(1-r^{2})x_{B}z,\\
\alpha_{c2}&=M^{2}_{B}[z-r^{2}(z-\overline{x}_{3})](x_{B}-\overline{x}_{3}\overline{\eta}),\\
\alpha_{d2}&=M^{2}_{B}[r^{2}(x_{3}-z)+z](x_{B}-x_{3}\overline{\eta}),\\
\alpha_{(e2, f2)}=\beta_{(g2, h2)}&=M^{2}_{B}[r^{2}(z-\overline{x}_{3})+\overline{z}](-\eta-x_{3}\overline{\eta}),\\
\beta_{e2}&=M^{2}_{B}[(1-r^{2})z-1],\\
\beta_{f2}&=M^{2}_{B}(1-r^{2}\overline{x}_{3})(-\eta-x_{3}\overline{\eta}),\\
\alpha_{g2}&=M^{2}_{B}\{[r^{2}(\overline{x}_{3}-z)+z](x_{B}-\overline{x}_{3}\overline{\eta})+r^{2}_{b}\},\\
\alpha_{h2}&=M^{2}_{B}[r^{2}(z-\overline{x}_{3})+\overline{z}](x_{B}-\eta-x_{3}\overline{\eta}),\\
\end{split}
\end{eqnarray}

The hard scales $t_{i}(i=a1,...h2)$, which are taken to remove the large logarithmic radiative corrections, are given by

\begin{eqnarray}
\begin{split}
t_{a1}&=\mathrm{Max}\left\{\sqrt{\left|\beta_{a1}\right|},1/b_{B}, 1/b_{3}\right\},t_{b1}=\mathrm{Max}\left\{\sqrt{\left|\beta_{b1}\right|},1/b_{B}, 1/b_{3}\right\},\\
t_{c1}&=\mathrm{Max}\left\{\sqrt{\left|\alpha_{c1}\right|}, \sqrt{\left|\beta_{c1}\right|},1/b_{B}, 1/b\right\},t_{d1}=\mathrm{Max}\left\{\sqrt{\left|\alpha_{d1}\right|}, \sqrt{\left|\beta_{d1}\right|},1/b_{B}, 1/b\right\},\\
t_{e1}&=\mathrm{Max}\left\{\sqrt{\left|\beta_{e1}\right|},1/b, 1/b_{3}\right\},t_{f1}=\mathrm{Max}\left\{\sqrt{\left|\beta_{f1}\right|},1/b, 1/b_{3}\right\},\\
t_{g1}&=\mathrm{Max}\left\{\sqrt{\left|\alpha_{g1}\right|}, \sqrt{\left|\beta_{g1}\right|},1/b_{B}, 1/b\right\},t_{h1}=\mathrm{Max}\left\{\sqrt{\left|\alpha_{h1}\right|}, \sqrt{\left|\beta_{h1}\right|},1/b_{B}, 1/b\right\},\\
t_{c2}&=\mathrm{Max}\left\{\sqrt{\left|\alpha_{c2}\right|}, \sqrt{\left|\beta_{c2}\right|},1/b_{B}, 1/b_{3}\right\},t_{d2}=\mathrm{Max}\left\{\sqrt{\left|\alpha_{d2}\right|}, \sqrt{\left|\beta_{d2}\right|},1/b_{B}, 1/b_{3}\right\},\\
t_{e2}&=\mathrm{Max}\left\{\sqrt{\left|\beta_{e2}\right|},1/b, 1/b_{3}\right\},t_{f2}=\mathrm{Max}\left\{\sqrt{\left|\beta_{f2}\right|},1/b, 1/b_{3}\right\},\\
t_{g2}&=\mathrm{Max}\left\{\sqrt{\left|\alpha_{g2}\right|}, \sqrt{\left|\beta_{g2}\right|},1/b_{B}, 1/b_{3}\right\},t_{h2}=\mathrm{Max}\left\{\sqrt{\left|\alpha_{h2}\right|}, \sqrt{\left|\beta_{h2}\right|},1/b_{B}, 1/b_{3}\right\}.\\
\end{split}
\end{eqnarray}

The Sudakov exponents are defined as

\begin{eqnarray}
\begin{split}
&S_{B}=s(x_{B}p^+_{1},b_{B})+\frac{5}{3}\int^{t}_{1/b_{B}}\frac{\emph{d}{\bar{\mu}}}{\bar{\mu}}\gamma_{q}(\alpha_{s}({\bar{\mu}})),\\
&S=s(zp^{+},b)+s(\bar{z}p^+,b)+2\int^{t}_{1/b}\frac{\emph{d}{\bar{\mu}}}{\bar{\mu}}\gamma_{q}(\alpha_{s}({\bar{\mu}})),\\
&S_{3}=s({x}_{3}p^{-}_{3},b_{3})+s(\bar{x}_{3}p^-_{3},b_{3})+2\int^{t}_{1/b_{3}}\frac{\emph{d}{\bar{\mu}}}{\bar{\mu}}\gamma_{q}(\alpha_{s}({\bar{\mu}})),
\end{split}
\end{eqnarray}
with the anomalous dimension of the quark $\gamma_{q}=-\alpha_{s}/\pi$, and $s(Q,b)$ is the Sudakov factor, which can be found in Ref.~\cite{H:2003fha}. Meanwhile, the threshold resummation factor $S_{t}(x)$ is taken from Ref.\cite{Kurimoto:2001xti},

\begin{equation}
S_{t}(x)=\frac{2^{1+2c}\Gamma(\frac{3}{2}+c)}{\sqrt{\pi}\Gamma(1+c)}[x(1-x)]^{c},
\end{equation}
with the parameter $c=0.3$.


\end{document}